\documentclass[twocolumn]{aastex631}

\newcommand{\msun}{{ M}_\odot}
\newcommand{\zsun}{{ Z}_\odot}
\usepackage{newtxtext,newtxmath}

\usepackage[normalem]{ulem}

\title{Metal-enriched Pair-instability supernovae: Effects of rotation}
\shortauthors{Umeda and Nagele}
\graphicspath{{./}{}}

\begin{document}

\title{Metal-enriched Pair-instability supernovae: Effects of rotation }

\author{Hideyuki Umeda}
\affiliation{Department of Astronomy, School of Science, The University of Tokyo, 7-3-1 Hongo, Bunkyo, Tokyo 113-0033, Japan}
\author{Chris Nagele}
\affiliation{Department of Astronomy, School of Science, The University of Tokyo, 7-3-1 Hongo, Bunkyo, Tokyo 113-0033, Japan}



\begin{abstract}

In this paper we revisit metal-enriched rotating pair instability supernovae (PISNe) models
for metallicities consistent with the Small Magellanic Cloud (SMC), 
the Large Magellanic Cloud (LMC) and 0.1$Z_\odot$. 
By calculating multiple models, we intend to clarify mass ranges and the ejected
 $^{56}$Ni masses from the PISNe, and mass loss histories for progenitors.
We find the choice of the Wolf-Rayet (WR) mass-loss rates are important and we adopt
the recently proposed rate of \citet{Sander2020MNRAS.499..873S}, which covers the mass ranges for PISNe progenitors.
We show that slow rotation lowers the PISN range due to the increase in core mass by rotational mixing.
On the other hand, if we assume typical rotation speed for observed OB stars, the mass loss increase
becomes more significant and the final stellar masses are smaller than non-rotating models.
As a result, typical mass range for bright SNe, with $^{56}$Ni mass more than 10${ M}_\odot$ 
for such fast rotating models are more than 400, 350 ${ M}_\odot$, for LMC and 0.1$Z_\odot$ metallicities,
respectively.
It is interesting that, unlike previous works, we find Oxygen rich progenitors for most cases.
This O-rich progenitor may be consistent with the recently identified PISN candidate SN2018ibb.
He-rich progenitors are seen only for relatively dim and metal poor ($Z \lesssim 0.1 Z_\odot$) PISNe.
We also discuss the black hole mass gap for metal-enriched PISNe, and show that 
the upperbound for the gap is lower than in the Pop III case.

\end{abstract}



\section{Introduction}
\label{introduction}

A pair-instability supernova (PISN) is a thermonuclear
 explosion of a massive Oxygen core
 induced by the creation of electron-positron pairs.
 After this mechanism was initially proposed \citep{Barkat1967PhRvL..18..379B,Rakavy1967ApJ...148..803R},
 detailed calculations confirmed the viability of the explosion \citep[e.g.][]{Umeda2002ApJ...565..385U,Heger2002ApJ...567..532H,takahashi2018}.
 These studies mostly focused on metal free Pop III stars because the mass range of PISNe is large (about $140 - 300 \;\msun $ for
 metal free stars) and stars massive enough to experience the pair instability are thought to be rare in the present universe because of large wind mass loss for metal enriched stars.
 Although there are some calculations for metal enriched stars
 \citep[e.g.][]{Langer2007A&A...475L..19L,Yoshida2011MNRAS.412L..78Y,Yusof2013MNRAS.433.1114Y,Kozyreva2014A&A...566A.146K,Whalen2014ApJ...797....9W}, all but two of these works (discussed below) assume that the
 main properties of PISNe, i.e., the explosion energy and the amount of
 $^{56}$Ni production, depend solely on CO-core mass and not on metallicity. \citet{Kozyreva2014A&A...566A.146K} and \citet{Whalen2014ApJ...797....9W}
 calculated hydrodynamics and nucleosynthesis for metal-enriched models ($Z \gtrsim 0.001$)
 during the exploding phase, but only two mass models are shown for each
 metal-enriched case.
 
 Observationally it is not clear if there has been any evidence for the
 existence of PISNe. Since the nucleosynthetic pattern is quite different from
 core collapse SNe, if an extremely metal poor star was formed from the gas of
 PISN ejecta, it should not be difficult to identify \citep[e.g.][]{Umeda2005ApJ...619..427U}. Recently, a single star with the signature of PISN enrichment was reported \citep{Xing2023Nature}, but no other metal poor stars with evidence of PISN enrichment have been found, although there is a suggestion that
 PISN abundance patterns might be hidden in very metal poor stars \citep{Aoki2014Sci...345..912A}.

A PISN can also be observed as a nearby supernova since in principle PISNe are possible
for $Z \lesssim Z_\odot/3$ \citep{Langer2007A&A...475L..19L}. The first suggestion for such a discovery
was reported as a Type I super luminous supernova (SLSN-I), SN2007bi \citep{Gal-Yam2009Natur.462..624G}.
However, spectroscopic models of PISNe were incompatible with the observations both in
the photospheric \citep{Dessart2012MNRAS.426L..76D} and nebular \citep{Jerkstrand2017ApJ...835...13J} phases.
Since then, magnetar models have more frequently been considered to explain SLSNe-I,
and PISNe models are rarely taken seriously. Nevertheless there remain some
SLSNe-I, such as PS1-14bj, PTF10nmn, OGLE14, and SN2020wnt,
which cannot be well explained by the magnetar models and these could well be PISNe. If these candidates are PISNe, the ejected $^{56}$Ni
mass falls in the range of $\sim 0.5 - 10M_\odot$ \citep{Gal-Yam2019ARA&A..57..305G}.
Very recently, there was a report that SN2018ibb might be the best PISN candidate discovered thus far \citep{Schulze2023arXiv230505796S}. SN2018ibb is a Hydrogen poor super-luminous SN
at $z=0.166$,
and if it is a PISN, the expected $^{56}$Ni mass is more than 30$\msun$. 
\citet{Schulze2023arXiv230505796S} also mention that there is a signature of interaction with oxygen-rich CSM.

  In \citet{Yoshida2011MNRAS.412L..78Y} we showed that if a nearby PISN with 
 metallicity $Z=0.004$ ejects more than a few solar masses $^{56}$Ni, the 
 initial stellar 
 mass should be more than 500$M_\odot$ for standard mass loss rates in the non-rotating
 models. On the other hand,  \citet{Yusof2013MNRAS.433.1114Y} found that
 rotating models can be such an PISN in the mass range 150-180 $M_\odot$ for
 SMC metallicity. In general, rotation increases core mass which may reduce the
 PISN mass range significantly. However there are some differences in the
 assumptions of these works and consequences of these differences is unclear. In this paper, we revisit metal-enriched rotating PISNs models to clarify
 model dependence of these PISNe. This is motivated by the paucity of models in the existing literature.

The remainder of the paper is organized as follows.
In Section~\ref{methods}, we summarize our numerical methods, including initial rotational profiles (\ref{sec:rot0}), mass loss rates (\ref{sec:massloss}), angular momentum transfer (\ref{sec:angular}), and finally the hydrodynamical calculations of the PISNe (\ref{sec:hydromethods}).
Section~\ref{results} shows mass loss histories (\ref{sec:masslossresults}), stellar evolution of rotating models (\ref{sec:rotation}), hydrodynamics (\ref{sec:hydroresults}), nucleosynthetic post processing (\ref{sec:nuc}) and 
some examples of peculiar results 
(\ref{sec:peculiar}). We then discuss our results in the context of previous work (Sec. \ref{sec:discussion}) before concluding in Sec. \ref{sec:summary}.

\section{Methods}
\label{methods}

 In this paper we calculate evolution of sub-solar metallicity stars
 in the PISN mass range for three metallicity cases: LMC metallicity
 ($Z = Z_\odot / 3$), SMC metallicity  ($Z = Z_\odot / 5$)
 and $Z = 0.1 Z_\odot $. 
 Here, $Z_\odot$ is the solar metallicity and we adopt $Z_\odot = 0.0141$.
 We note that the SMC and LMC metallicities mentioned above roughly
 corresponds to the lower metallicity bounds for these galaxies.

For this study, we use the HOSHI (HOngo Stellar Hydrodynamics Investigator) code 
described e.g. in \citet{takahashi2018} and \citet{Luo2022ApJ...927..115L}
for the stellar evolution.
We follow the nuclear burning using a nuclear reaction network of 153 species of nuclei \citep{takahashi2018}.
Nuclear reaction rates are taken from the JINA reaclib database v1 \citep{Cyburt2010ApJS..189..240C}, except for the $^{12}C(\alpha,\gamma)^{16}O$ rate which is taken to be 1.5 times the value given in \citet{caughlan1988}. 
In this paper we adopt the overshooting parameter $f_{\rm ov}$=0.01 as same as in
\citep{takahashi2018}, which is called the `M' model
in \citet{Luo2022ApJ...927..115L}.
Then we use the hydrodynamical code described in \citet{takahashi2016} and \citet{Nagele2022arXiv220510493N}
for the collapse and PISN explosions after the model approaches  
the electron-positron pair-creation instability region. 

Calculation methods for stellar evolution are mostly the same 
as in our previous work, but there are differences in the setting of initial models, 
and the treatments in the mass loss rates
as explained in Sec 2.2.
Previously we started stellar evolution from a zero age main sequence (ZAMS) model and
set the initial rotation speed for that ZAMS model. However, the definition of
ZAMS is somewhat uncertain and it is thus difficult to compare with other works
 which may adopt a slightly different definition for ZAMS. Usually this small
 difference does not matter, but in this work initial rotation speed is 
critical, so we adopt a more concrete definition.

\subsection{Initial rotation}
\label{sec:rot0}

\begin{figure}
    \centering
    \includegraphics[width=1.0\linewidth]{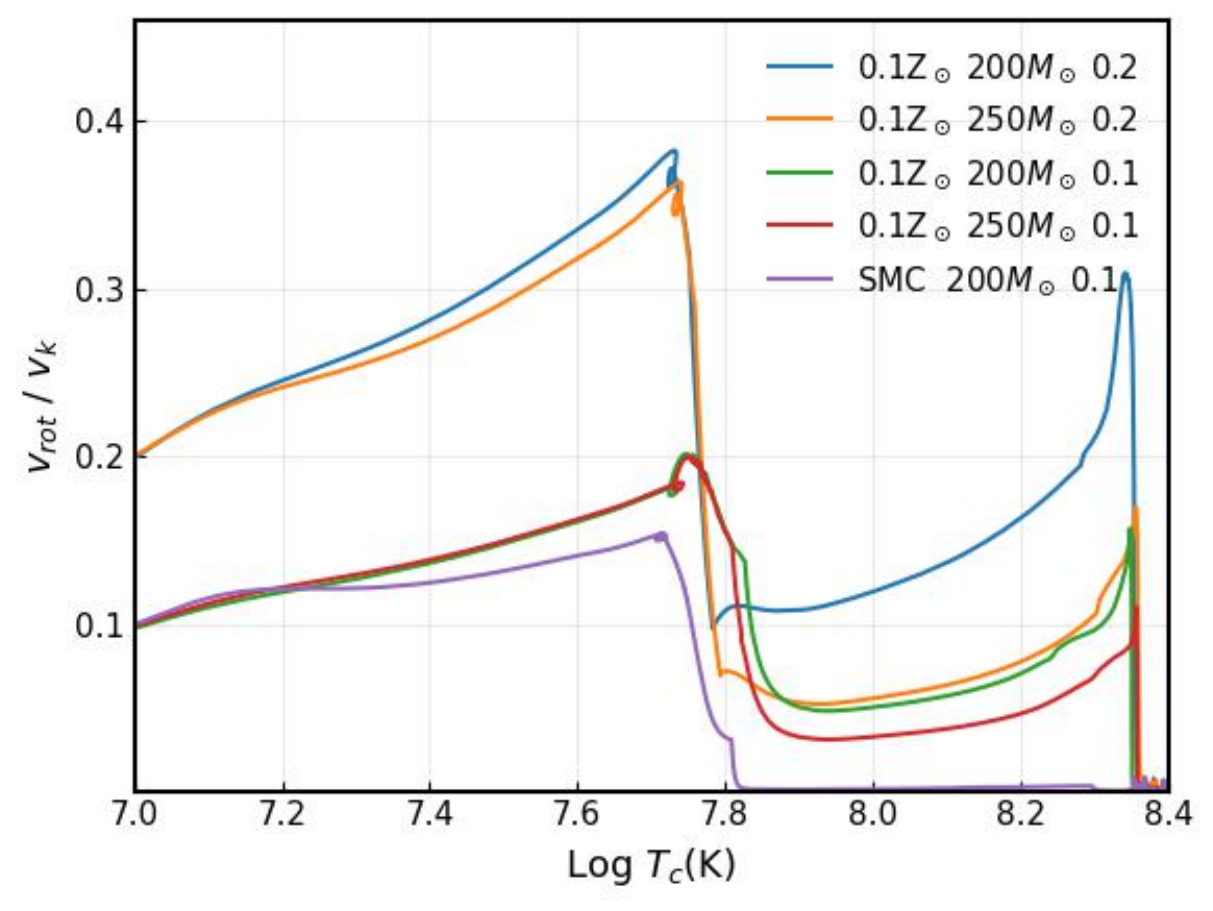}    
    \caption{Evolution of $v_{\rm rot}/v_{\rm K}$ for some models 
    as a function of Log $T_{\rm c}$ up to the early He burning stage. 
   }
    \label{fig:vrot_Tc_comp}
\end{figure}

We now describe the determination of initial rotation in this work.
First we construct an expanded pre-mainsequence model which has 
central temperature below $\log T_{\rm c} < 7.0$ (K). Then we begin the  stellar evolution calculation, and when the central temperature reaches $\log T_{\rm c} 
= 7.0$ (K),
we set the initial rotational velocity. We assume rigid rotation initially and
the ratio of the surface velocity, $v_{\rm i}$, 
to the Keplerian velocity, $v_{\rm K} \equiv \sqrt{GM/R}$,
is taken as a parameter. As shown below $v_{\rm i}/v_{\rm K} = 0.1$
corresponds to relatively slow rotation, and 0.2 to fast-rotation.

Figure \ref{fig:vrot_Tc_comp} shows typical evolution for the ratio of the
surface velocity to the Keplerian velocity, $v_{\rm rot}/v_{\rm k} $.
These lines are for 0.1 $Z_\odot$, $M=$ 200 and 250 $M_\odot$
with $v_{\rm i}/v_{\rm k} = 0.1$ and $0.2$, and SMC metallicity, $M=$ 200 $M_\odot$
with $v_{\rm i}/v_{\rm k} = 0.1$.
As shown in this figure, $v_{\rm rot}/v_{\rm k}$ 
increases with the initial stellar shrinkage, and has a maximum
around ZAMS, where $\log T_c$ (K) $\sim 7.7$. These maximal values are 
shown in Table 1 as $(v_{\rm rot}/v_{\rm k})_{\rm max} $.
We note that the SMC model has lower $(v_{\rm rot}/v_{\rm k})_{\rm max} $
than the 0.1 $Z_\odot$ model with the same $M$ and $v_{\rm i}/v_{\rm k}$ due to
slightly larger mass loss before ZAMS.

This figure also shows that $(v_{\rm rot}/v_{\rm k})_{\rm max} $ is about
0.2 for $v_{\rm i}/v_{\rm k} = 0.1$ and 0.3-0.4 for $v_{\rm i}/v_{\rm k} = 0.2$.
Since the typical $v_{\rm rot}/v_{\rm k}$ of massive main sequence stars is
about 0.3-0.4 \citep[e.g.][]{Georgy2012A&A...542A..29G}, 
we may say that $v_{\rm i}/v_{\rm k} = 0.2$ models rotate rapidly, and
$v_{\rm i}/v_{\rm k} = 0.1$ models are relatively slowly rotating, though 
we do not have any observational evidences that very massive stars considered
in this paper rotate as fast as observed massive stars.

\subsection{Mass loss rates}
\label{sec:massloss}

   We will find that the mass loss rates, especially the rate
  for Hydrogen poor stars, are important for metal-enriched PISNe. Therefore, 
  in this subsection we describe in detail the mass loss rates we use.
  For Hydrogen-rich stars with the surface Hydrogen mass fraction, X(H), higher than 0.5,
  we use a rate by \citet{Vink2001A&A...369..574V} if the effective temperature is higher than $\log T_{\rm eff}$= 4.05. We use a different rate
  for Hydrogen-rich (X(H)$>0.3$) Yellow and Red supergiants with $\log T_{\rm{eff}} \; < 3.7$ 
  as in \citet{Sylvester1998MNRAS.301.1083S,vanLoon1999A&A...351..559V,Crowther2000A&A...356..191C,Ekstrom2012A&A...537A.146E}.
  If $\log T_{\rm eff} > 4.05$ and X(H) $< 0.3$, we use the rate for Wolf-Rayet (WR) stars. 
  If the parameter space is outside the region mentioned above, we use 
  \citet{deJager1988A&AS...72..259D}. We note that as shown in \citet{Yusof2013MNRAS.433.1114Y},
  the very massive stars considered in this paper typically have large 
  $\log T_{\rm{eff}}$, say $ >4 $, throughout the evolution, so we are most often using the  
  \citet{Vink2001A&A...369..574V} rate for Hydrogen-rich stars.

 Previously, e.g. in \citet{Yoshida2019} and later work using the HOSHI code,
  we have used \citet{Nugis2000A&A...360..227N} for hydrogen poor stars, and 
  used \citet{deJager1988A&AS...72..259D} outside the region of  
   \citet{Nugis2000A&A...360..227N}. 
   However, we have found that the mass loss rate
   sometime becomes small in this approach and we had sometimes significantly larger final
   mass than other groups' work. Therefore, we decided to replace the Wolf-Rayet
   mass loss rate with more commonly used ones from recent similar studies.
  Specifically, we tried two types. One type is the rate used in \citet{Aguilera-Dena2018ApJ...858..115A,Aguilera-Dena2020ApJ...901..114A} and \citet{Yoon2017MNRAS.470.3970Y} (hereafter Y2017 rate).
  The other is a theoretical rate given in \citet{Sander2020MNRAS.499..873S} (SV2020).
   After implementing these rates, we had roughly consistent results with other works.
   For example, the final mass of solar metallicity $\sim 50 M_\odot$ models are
   about ten solar masses.

  There are some differences in the two WR mass loss rates. As shown in 
  SV2020, their rate is typically smaller than the Y2017 rate. 
  There is also a notable difference in the metallicity
  dependence. The rate by SV2020 decreases rapidly
  toward lower metallicity compared with other rates in the literature. 
  Because of these differences we found that the PISN mass range is significantly
  larger if Y2017 rate is used. Since the Y2017 rate is determined mainly to
  explain observed properties of WR stars, it is not clear if we can extrapolate it
  to the PISN region. On the other hand, SV2020 rate can be applicable up to
  500 $\msun$ and metallicity down to 0.02 $Z_\odot$. Therefore, we use the SV2020
  rate in this work, though we emphasize that the mass loss rate
  is quite uncertain. More specifically, for the WR stars, we use the expression in the
  equation (28) of SV2020 and parameters given in their equations from (23) to (32).

 When a star is rotating rapidly,
 mass loss is enhanced due to the nearly-critical 
 rotation at the surface (the $\Omega - \Gamma$
 limit, \citealt{Langer1998A&A...329..551L,Maeder2000A&A...361..159M}). 
 In order to include this effect, we consider the following expression
 for the mass loss rate
 according to \citet{Yoon2012A&A...542A.113Y}: 
 \begin{equation}
    \dot M= - \min \left[|\dot M (v_{\rm rot}=0)|\times 
    \left( 1-\frac{v_{\rm rot}}{v_{\rm crit}}\right)^{-0.43}, 
    0.3 \frac{M}{\tau_{\rm KH}} \right],
\end{equation}
where $v_{\rm rot}$, $v_{\rm crit} \equiv \sqrt{GM(1-L/L_{\rm Edd)}/R}$, $\tau_{\rm KH}, L_{\rm edd}$ are the surface rotation velocity, the critical rotation velocity, the Kelvin-Helmholtz timescale, and the Eddington luminosity, respectively. There is an ambiguity in the definition of the Kelvin-Helmholtz timescale $\sim GM^2/RL$, so we introduce a parameter of $O(1)$, $f_{\rm KH}$, such that 
$\tau_{\rm KH} = f_{\rm KH} GM^2/RL$. In this paper we adopt $f_{\rm KH}=0.5$ \citep{Kippenhahn1990sse..book.....K}.

 As shown above, we set an upper limit on the magnitude of the mass loss rate
 and refer to it as $\dot M_{\rm up} \equiv 0.3M/\tau_{\rm KH} $. 
 We refer to the magnitude of the non-critical rate as 
 $\dot M_{\rm rot} \equiv |\dot M (v_{\rm rot}=0)|\times ( 1- v_{\rm rot}/v_{\rm crit})^{-0.43}$.

 In general, the rotation induced mass loss rate increases when a star is shrinking, because
 both the $v_{\rm rot}$ and $L$ tend to increase. The reader might think that in these
 stages, $\dot M_{\rm rot}$ increases gradually and when it reaches $\dot M_{\rm up}$,
 we replace it with the upper limit. However, this is not the typical realization.
 Since we cannot adopt very short time steps to follow the star over evolutionary timescales, in a realistic
 calculation, the ratio $v_{\rm rot}/v_{\rm crit}$ often approaches and then exceeds 1. We use $\dot M_{\rm up}$ if $v_{\rm rot}/v_{\rm crit}$ exceeds 1.
 In the usual calculations, $\dot M_{\rm rot} \ll \dot M_{\rm up}$ even if 
 $v_{\rm rot}/v_{\rm crit}$ is very close to 1. Therefore, in practice
 we use $\dot M_{\rm rot}$ for $v_{\rm rot}/v_{\rm crit} < 1$
and use $\dot M_{\rm up}$ for $v_{\rm rot}/v_{\rm crit} \geq 1$.

 This inclusion of the upper limit for mass-loss rate is different from 
 recent related work,
 \citet{Aguilera-Dena2018ApJ...858..115A,Aguilera-Dena2020ApJ...901..114A}, 
 where they instead
 set the upper limit to the value $v_{\rm rot}/v_{\rm crit}$ to be 0.98.
 This difference may affect the results for rapidly rotating cases;
 however, the difference is not important in this paper, since it turns 
 out that all the models shown in this paper end up as slow rotators
 because of large mass loss during the late He-burning and WR stages.

\begin{table*}
	\centering
	\caption{The initial stellar mass range for becoming PISNe 
 and the (most) abundant elements at the surface of the final models.}
	\label{tab:massrange}

	\begin{tabular}{cccc}  
		    \hline \hline
    Metallicity & $v_{\rm i}/v_{\rm k}$ & mass range ($\msun$) & Surface abundance \\
    		 \hline
      $Z_\odot$/3 (LMC) & 0 & 320 $\sim $ 475 & O-rich  \\
                  & 0.1 &  $  \gtrsim 400 $ & O-rich  \\
                  \hline
      $Z_\odot$/5 (SMC) & 0 & 300 $\sim $ 390 & O-rich  \\
            & 0.1 & 240 $\sim $ 350 & O-rich  \\
            & 0.2 &    $ \gtrsim 360$   & O-rich  \\
            \hline
      0.1 $Z_\odot$ & 0 & 140* $\sim $ 330 & He/H $\rightarrow$ O ($M\geq$ 180) \\
            & 0.1 & 140 $\sim $ 250 & He/H $\rightarrow$ O ($M\geq$ 200)  \\
            & 0.2 & 220 $\sim $ 410 & O-rich  \\
            & 0.3 & 240 $\sim $ 430* & O-rich  \\      
 
		\hline \hline
	\end{tabular}
\end{table*}

\subsection{Angular momentum transfer}
\label{sec:angular}

 In the HOSHI code, the diffusion approximation is applied for the transportation of angular momentum, 
similar to the codes in \citet{Yoon2005A&A...443..643Y,Woosley2006ApJ...637..914W} which were used to calculate
progenitor models of long gamma-ray bursts (GRBs). As in these works, we assume a 
magnetic model and the Tayler-Spruit dynamo (TS dynamo, \citealt{Spruit2002A&A...381..923S}), is applied.
With the TS dynamo, angular momentum is transferred efficiently when differential rotation
exists. 
For the inclusion of the TS dynamo effect, we follow the method in \citet{heger2005} as described in
\citet{Takahashi2014ApJ...794...40T}.
Because of this effect, the stellar cores are relatively slowly rotating after the He burning stages. It is then difficult to produce the fast rotating cores that are 
necessary for GRB progenitors \citep{heger2000} unless
chemically homogeneous evolution (CHE) occurs (\citet{Yoon2005A&A...443..643Y}). 
Since we have not applied our code to CHE models and also adopt a different
WR mass loss rate from previous works, we tested whether we can produce
CHE stars. We confirmed that CHE stars with a fast rotating
Fe core can be obtained with our code for metal poor, i.e., below 0.1 Z$_\odot$, 
$\sim 30-50 \msun$ stars. In this paper, however, we focus on how rotation affects the PISN explosion mass range.

\subsection{Hydrodynamical and nucleosynthesis calculation}
\label{sec:hydromethods}

 In this paper, we aim to calculate explosion and nucleosynthesis for pair-instabillity
 supernovae. Such calculations were done in \citet{takahashi2018} for Pop III PISNe
 and we use basically the same method. We calculate stellar evolution with the HOSHI code, until
 the central temperature reaches around Log Tc =9.2 and then we switch to the hydrodynamical
 code. The HOSHI code includes the acceleration term
 in the equation of motion and in principle could calculate hydrodynamics. However, since this is
 a Henyey type stellar evolution code, energy conservation is not as good as in a hydrodynamical
 code and not suitable for PISN simulations. This is the reason we switch to a hydrodynamical code.

The hydrodynamics code is a one dimensional Lagrangian general relativistic hydrodynamics code \citep{yamada1997,sumiyoshi2005,takahashi2016}. The code includes energy changes due to nuclear reactions and neutrino cooling, and we use the same 153 isotope network as in HOSHI. In this paper, we use 255 radial meshes and the HOSHI models are mapped to this grid using the same procedure as in previous works \citep[e.g.][]{takahashi2018} while the numerical settings are identical to those used in \citet{Nagele2023arXiv230101941N}. After shock breakout, the timesteps become large and we terminate the simulation. For one LMC model (100 $\msun$), a pulsation occurs instead of an explosion, and for this model we terminate the calculation after shock breakout, as we are not overly interested in pulsations in this paper.

After the hydrodynamics calculation is completed, we post process the nucleosynthesis with a network consisting of 300 nuclei, as in \citet{takahashi2018}. The $^{56}$Ni masses reported in this paper are from the post processing, and are slightly lower than the values obtained from hydrodynamics. After post processing the hydrodynamics, additional post processing is carried out with fixed temperature and density until the composition is fully decayed. The reported elemental yields use the abundances at the end of this additional post processing.

\section{Results}
\label{results}

\subsection{PISN Mass Range }
\label{sec:massrange}

 In Table \ref{tab:massrange}, we show the initial stellar mass range 
 for the models which become PISNe and the (most) abundant elements at the surface
 of the models in our results.
 For example, He/H $\rightarrow$ O ($M\geq$ 180) means that the
 surface abundance is He and H-rich for lower masses and O-rich 
 above $M = 180 \msun$. In these models, the mass fractions of 
 He and H are about 0.8 and 0.2, respectively.

  In general, the PISN range is higher for larger metallicity since the
  mass loss rate is larger. 
  The effects of stellar rotation on the PISN range can be summarized as follows.
  Slow rotation lowers the range slightly since the He-core becomes larger due to
  rotational mixing. However, in this parameter range, faster rotation
  increases the mass range since the stars become WR-stars earlier because of
  rotation induced mass loss, meaning that the final mass becomes smaller 
  due to the efficient mass-loss during the WR stages.
  For much lower metallicity where the mass loss of WR stars is small,
  we may find CHE stars becoming a PISN. However, we have not found such a
  case in this metallicity range.

There are a few previous studies for metal-enriched PISNe.
In \citet{Whalen2014ApJ...797....9W}, using the GENEVA stellar evolution code \citep{Hirschi2004A&A...425..649H,Eggenberger2008Ap&SS.316...43E},
they showed that 150 and 200 $\msun$ with 0.1 $Z_\odot$ 
and 500 $\msun$ with 0.3 $Z_\odot$ become PISNe. Our results are broadly
similar though there are some specific differences.
Their models assume relatively fast rotation, which corresponds to
$v_{\rm i}/v_{\rm k} \gtrsim 0.2$ in our definition, and 
our models yield O-rich PISNe within a higher mass range. 
This mainly originates from the difference in the mass loss
rate, especially in the WR stages. 

\citet{Yusof2013MNRAS.433.1114Y} also provided a PISN mass range using the GENEVA code. 
Their non-rotaing LMC model became a PISN for $M \gtrsim 300 \msun$.
This is roughly consistent with our results, as their rotating models also have a higher
mass range. Their results for rotating SMC models are
quite different from ours. They show that 150 and 200 $\msun$ models
become PISNe with large amounts of $^{56}$Ni production, while 
for us these models do not become PISNe. Our models require lower metallicity
and slower rotation to make such energetic PISNe.

\citet{Yoshida2011MNRAS.412L..78Y} also showed the range for metallicity
 $Z=0.004$ non-rotating models. 
 Compared with our non-rotating SMC models, their lower bound is similar
 but their upper bound is much larger (more than $500 \msun$). In addition they
 found only He-rich models. These differences also originate from the chosen WR mass-loss rates.
 In the followings we describe in more detail how we got these results.
 
\begin{table*}
	\centering
	\caption{Summary table for models denoted by the initial mass ([$\msun$]), metallicity, and initial rotation. The remaining columns show maximum rotation, final mass, final CO core mass (in units of $\msun$), mass loss rate at $\log T_{\rm c}$ =9.0 (K) 
   ([$\msun$/year]), surface composition and the mass fraction of it in ( ), explosion energy, maximum central temperature during the explosion, and $^{56}$Ni mass as determined by the 300 isotope post processing. $^\dagger$This model is incorrectly denoted as an explosion in the published version. This table has the correct information.}

	\label{tab:hydro}
	\begin{tabular}{cccccccccccc} 
		    \hline \hline
    		$M_{\rm i}$ & $Z$ & $v_{\rm i}/v_{\rm k} $ &$(v_{\rm rot}/v_{\rm k})_{\rm max} $ &  $M_{\rm fin}$ &  $M_{\rm CO}$ & $ |\dot M_{\rm 9.0}| $ & surface & fate & $E_{\rm exp}$ [$10^{52}$ ergs] & Max Log $T_c$  & $^{56}$Ni [$\msun$] \\
    		\hline
310 & LMC & 0 & 0& 67.4 & 67.4 & 8.6E-4 & O(0.84) & PPISN & --- & 9.47 &  0 \\
320 & LMC & 0 & 0& 67.4 & 67.4 & 8.6E-4 & O(0.84) & PISN & 1.08 & 9.57 &  0.33 \\
330$^\dagger$ & LMC & 0 & 0& 63.4 & 63.4 & 8.1E-4 & O(0.84) & CC$^\dagger$ & 0    & --- & 0 \\
340 & LMC & 0 & 0& 65.6 & 65.6 & 7.3E-4 & O(0.84) & PPISN & --- & 9.57 & 0 \\
350 & LMC & 0 & 0& 63.2 & 63.2 & 9.0E-4 & O(0.84) & PISN & 0.98 & 9.57 & 0.28 \\
360 & LMC & 0 & 0&69.6 & 69.6 & 8.9E-4 & O(0.84) & PISN & 1.33 & 9.59 & 0.51 \\
370 & LMC & 0 & 0&74.0 & 74.0 & 9.4E-4 & O(0.84) & PISN & 1.87 & 9.61 & 1.09\\
380 & LMC & 0 & 0&73.1 & 73.1 & 9.3E-4 & O(0.84) & PISN & 1.77 & 9.61 & 0.95\\
390 & LMC & 0 & 0&75.7 & 75.7 & 9.8E-4 & O(0.84) & PISN & 2.02 & 9.62 & 1.36\\
400 & LMC & 0 & 0&78.1 & 78.1 & 1.0E-3 & O(0.84) & PISN & 2.28 & 9.63 & 1.98\\
450 & LMC & 0 & 0&97.8 & 97.8 & 1.2E-3 & O(0.83) & PISN & 4.62 & 9.63 & 16.3\\
460 & LMC & 0 & 0&102.2 & 102.2& 1.3E-3 & O(0.83)& PISN & 5.32 & 9.74 & 22.1\\
470 & LMC & 0 & 0&104.5 & 104.5& 1.3E-3 & O(0.83)& PISN & 5.63 & 9.76 & 25.2\\
475 & LMC & 0 & 0&109.0 & 109.0& 1.3E-3 & O(0.83)& PISN & 6.28 & 9.79 & 32.4\\
480 & LMC & 0 & 0&119.7 & 119.7& 1.5E-3 & O(0.82)& CC   & 0    & --- & 0 \\
\hline
390 & LMC & 0.1 & 0.16 &70.4 & 70.4 & 7.5E-4 & O(0.84) & PPISN & --- & 9.45 & 0 \\
400 & LMC & 0.1 & 0.16 &71.3 & 71.3 & 9.1E-4 & O(0.81) & PISN & 1.55 & 9.60 & 0.70 \\
450 & LMC & 0.1 & 0.16 &71.3 & 71.3 & 9.1E-4 & O(0.84) & PISN & 1.55 & 9.60 &  0.70    \\
500 & LMC & 0.1 & 0.16 &75.8 & 75.8 & 9.7E-4 & O(0.84) & PISN & 2.03 & 9.62 &  1.4    \\
\hline
290 & SMC & 0  & 0&63.2 & 63.2 & 7.2E-4 & O(0.83) & PPISN & --- & 9.49 & 0 \\
300 & SMC & 0  &0& 69.2 & 69.2 & 8.2E-4 & O(0.84) & PISN & 1.28 & 9.58 & 0.47 \\
330 & SMC & 0  & 0&77.0 & 77.0 & 9.0E-4 & O(0.84) & PISN & 2.06 & 9.63 & 1.71 \\
340 & SMC & 0  & 0&83.2 & 83.2 & 9.8E-4 & O(0.84) & PISN & 3.00 & 9.65 & 4.69 \\
350 & SMC & 0  & 0&86.8 & 86.8 & 1.0E-3 & O(0.84) & PISN & 3.25 & 9.66 & 6.04 \\
360 & SMC & 0  & 0&86.9 & 86.9 & 1.0E-3 & O(0.84) & PISN & 3.28 & 9.66 & 6.18 \\
370 & SMC & 0  & 0&95.9 & 95.9 & 1.1E-3 & O(0.83) & PISN & 4.44 & 9.71 & 14.0 \\
380 & SMC & 0  & 0&108.6 & 108.6 & 1.2E-3 & O(0.83) & PISN & 6.07 & 9.81 & 30.0 \\
380B & SMC & 0 & 0&113.6 & 113.6 & 1.2E-3 & O(0.83) & PISN & 6.95 & 9.83 & 39.1 \\
390 & SMC & 0  & 0&96.3 & 96.3 & 1.1E-3 & O(0.83) & PISN & 4.47 & 9.71 & 14.3 \\
400 & SMC & 0  & 0&116.7 & 116.7 & 1.3E-3 & O(0.83) & CC & 0 & --- & 0 \\
\hline
230 & SMC & 0.1 & 0.16 &54.6 & 54.6 & 6.3E-4 & O(0.83) & PPISN  & --- & 9.45 & 0 \\
240 & SMC & 0.1 & 0.16 &70.1 & 70.1 & 7.6E-4 & O(0.84) & PISN & 0.34 & 9.57 & 0.12 \\
250 & SMC & 0.1 & 0.16 &70.1 & 70.1 & 8.2E-4 & O(0.75) & PISN & 1.36 & 9.59  & 0.52  \\
260 & SMC & 0.1 & 0.16 &69.1 & 69.1 & 8.1E-4 & O(0.84) & PISN & 1.24 & 9.58 & 0.44 \\
280 & SMC & 0.1 & 0.16 &72.0 & 72.0 & 8.4E-4 & O(0.84) & PISN & 1.59 & 9.60 & 0.73 \\
300 & SMC & 0.1 & 0.17 &78.4 & 78.4 & 9.2E-4 & O(0.84) & PISN & 2.27 & 9.63 & 1.83 \\
320 & SMC & 0.1 & 0.16 &88.9 & 88.9 & 1.0E-3 & O(0.84) & PISN & 3.51 & 9.67 &  7.5 \\
330 & SMC & 0.1 & 0.17 &111.0 & 111.0 & 1.2E-3 & O(0.83) & CC & 0 & --- & 0    \\
331 & SMC & 0.1 & 0.17 &94.7 & 94.7 & 1.0E-3 & O(0.84) & PISN & 3.85 & 9.72 &  10.8   \\
340 & SMC & 0.1 & 0.17 &117.8 & 117.8 & 1.3E-3 & O(0.82) & CC & 0 & --- & 0 \\
350 & SMC & 0.1 & 0.17 &113.8 & 113.8 & 1.5E-3 & O(0.83) & PISN & 6.98 & 9.83 & 39.5 \\
		\hline \hline
	\end{tabular}
\end{table*}

\begin{figure}
    \centering        
    \includegraphics[width=1.0\linewidth]{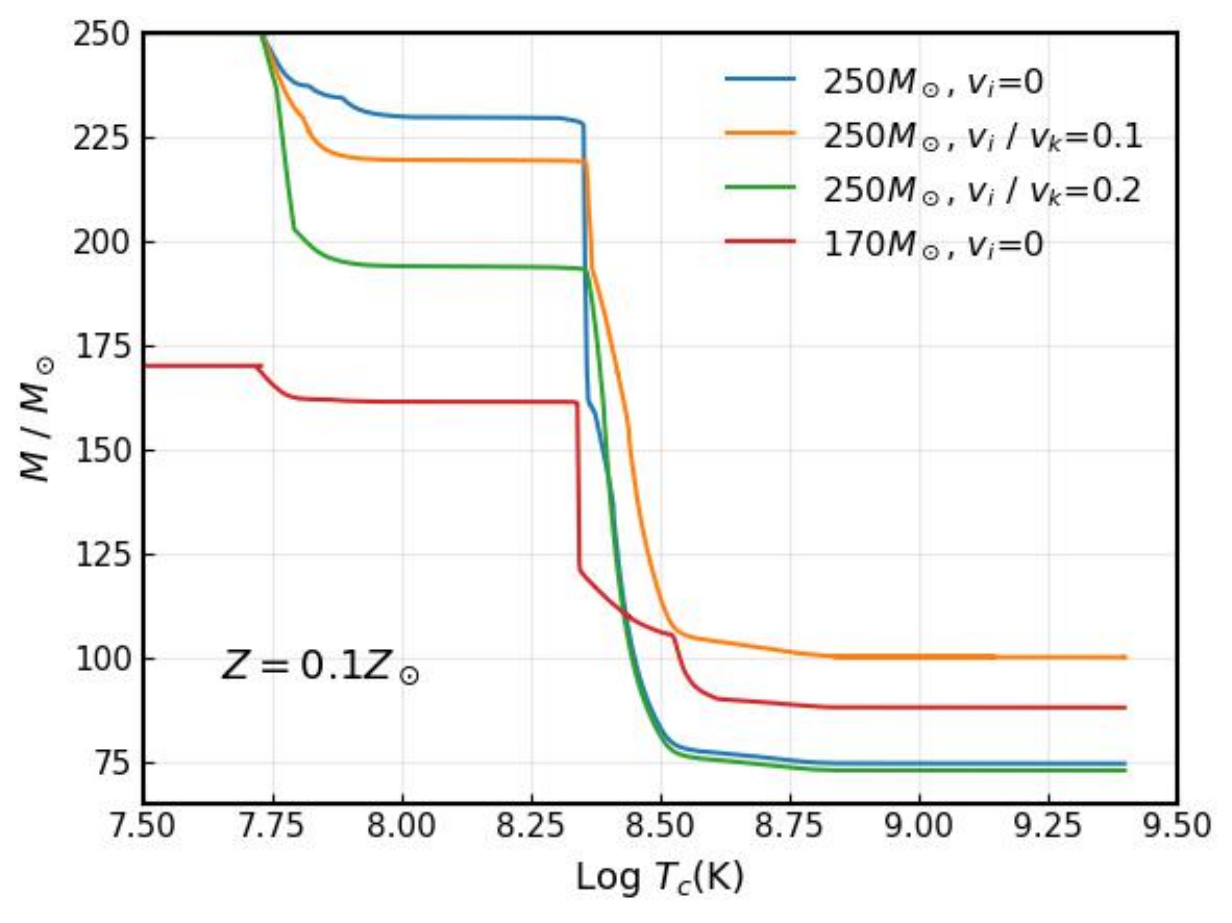}    
    \caption{Total mass evolution as a function of $\log T_{\rm c}$
    for the selected models in $Z=0.1Z_\odot$.}
    \label{fig:M_Tc}
\end{figure}

\begin{figure}
    \centering        
    \includegraphics[width=1.0\linewidth]{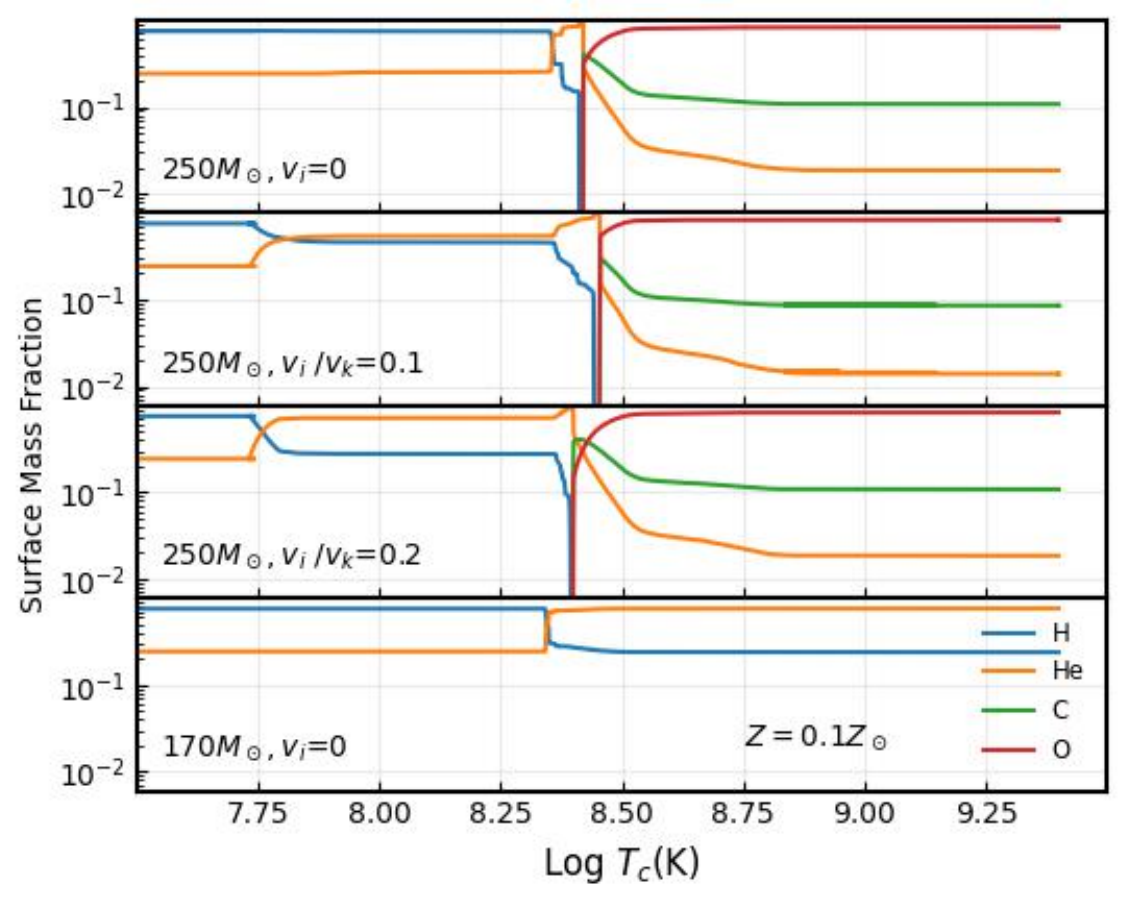}    
    \caption{Surface abundance evolution for the models in Figure \ref{fig:M_Tc}.}
    \label{fig:X_Tc}
\end{figure}

\begin{table*}
	\centering
	\caption{Same as Table 2, but for $Z=0.1 Z_\odot$ models.}
	\label{tab:hydro1}
	\begin{tabular}{cccccccccccc} 
		    \hline \hline
    		$M_{\rm i}$ & $Z$ & $v_{\rm i}/v_{\rm k} $ & $(v_{\rm rot}/v_{\rm k})_{\rm max} $ & $M_{\rm fin}$ &  $M_{\rm CO }$ & $ |\dot M_{\rm 9.0}| $ & surface & fate & $E_{\rm exp}$ [$10^{52}$ ergs] & Max Log $T_c$  & $^{56}$Ni [$\msun$] \\
    		\hline
130 & 0.1 $Z_\odot$ & 0 & 0 &67.7 & 57.8 & 6.2E-3 & He(0.74) & PPISN & --- & 9.43 & 0 \\
140 & 0.1 $Z_\odot$ & 0 & 0 &78.8 & 62.7 & 2.6E-3 & He(0.77) & PISN & 0.49 & 9.52 & 0.39 \\
150 & 0.1 $Z_\odot$ & 0 & 0 &87.9 & 68.3 & 4.6E-3 & He(0.75) & PISN & 0.94 & 9.56 & 0.173 \\
160 & 0.1 $Z_\odot$ & 0 & 0 &61.6 & 61.6 & 9.0E-4 & O(0.81) &  PISN  & 0.31 & 9.52  & 0.011 \\
170 & 0.1 $Z_\odot$  & 0 & 0 &87.9 & 78.2 & 8.9E-4 & He(0.76) & PISN & 1.33 & 9.59 & 1.01 \\
180 & 0.1 $Z_\odot$  & 0 & 0 &76.1 & 76.1 & 9.4E-4 & O(0.82) & PISN & 1.87 & 9.61 & 0.93\\
200 & 0.1 $Z_\odot$  & 0 & 0 &75.3 & 75.3 & 9.3E-4 & O(0.83) & CC   & 0 & --- & 0\\
210 & 0.1 $Z_\odot$  & 0 & 0 &76.1 & 76.1 & 7.8E-4 & O(0.83) & PISN & 1.93 & 9.61 & 1.04  \\
250 & 0.1 $Z_\odot$  & 0 & 0 &74.0 & 74.0 & 7.6E-4 & O(0.83) & PISN & 1.79 & 9.61 & 0.82\\
300 & 0.1 $Z_\odot$  & 0 & 0 &94.8 & 94.8 & 9.7E-4 & O(0.84) & PISN & 4.23 & 9.69 & 11.9\\
320 & 0.1 $Z_\odot$  & 0 & 0 &106.9 & 106.9 & 1.1E-3 & O(0.83) & PISN & 5.77 & 9.78 & 26.2\\
330 & 0.1 $Z_\odot$  & 0 & 0 &112.5 & 112.5 & 1.2E-3 & O(0.83) & PISN & 6.73 & 9.80 & 36.0 \\
335 & 0.1 $Z_\odot$  & 0 & 0 &119.2 & 119.2 & 1.2E-3 & O(0.83) & CC & 0 & --- & 0\\
\hline
130 & 0.1 $Z_\odot$ & 0.1 & 0.22 &86.5 & 66.5 & 1.6E-4 & He(0.81) & PPISN  & --- & 9.43 & 0 \\
140 & 0.1 $Z_\odot$ & 0.1 & 0.21 &96.1 & 71.0 & 7.1E-4 & He(0.79) &  PISN & 1.43 & 9.58  & 0.39 \\
150 & 0.1 $Z_\odot$ & 0.1 & 0.21 &95.9 & 80.0 & 9.6E-4 & He(0.82) &  PISN & 2.53 & 9.61 & 1.29  \\
170 & 0.1 $Z_\odot$ & 0.1 & 0.20 &108.8 & 89.2 & 3.6E-4 & He(0.87) &  CC  & 0 & --- & 0  \\
180 & 0.1 $Z_\odot$ & 0.1 & 0.20 &119.5 & 97.0 & 4.2E-4 & He(0.83) & PISN & 4.72  & 9.68& 9.87 \\
190 & 0.1 $Z_\odot$ & 0.1 & 0.18 &131.0 & 102.5 & 2.7E-4 & He(0.81) & PISN & 5.11  & 9.72  & 14.0 \\
200 & 0.1 $Z_\odot$ & 0.1 & 0.20 &101.2 & 101.2 & 1.0E-3 & O(0.83) & PISN & 4.89  & 9.73  & 16.9 \\
210 & 0.1 $Z_\odot$ & 0.1 & 0.21 &77.7 & 77.7 & 7.8E-4 & O(0.83) & PISN & 1.84  & 9.62  & 1.16 \\
220 & 0.1 $Z_\odot$ & 0.1 & 0.21 &88.4 & 88.4 & 9.3E-4 & O(0.83) & PISN & 3.74  & 9.68  & 8.32 \\
230 & 0.1 $Z_\odot$ & 0.1 & 0.22 &106.8 & 106.8 & 1.1E-3 & O(0.83) & PISN & 5.84  & 9.75  & 25.2 \\
230B & 0.1 $Z_\odot$ & 0.1 & 0.22 &118.3 & 118.3 & 1.2E-3 & O(0.79) & PISN & 7.00  & 9.80  & 37.5 \\  
240 & 0.1 $Z_\odot$ & 0.1 & 0.20 &107.3 & 107.3 & 9.0E-4 & O(0.83) & CC & 0  & ---  & 0  \\
250 & 0.1 $Z_\odot$ & 0.1 & 0.20 &100.0 & 100.0 & 8.9E-4 & O(0.83) & PISN & 4.04  & 9.78  &  13.0\\
255 & 0.1 $Z_\odot$ & 0.1 & 0.20 &149.4 & 149.4 & 1.4E-3 & He(1.00) & CC & 0  & ---  & 0 \\
\hline
210 & 0.1 $Z_\odot$ & 0.2 & 0.39 &66.0 & 66.0 & 6.3E-4 & O(0.83) &  PPISN  & --- & 9.56 & 0 \\
220 & 0.1 $Z_\odot$ & 0.2 & 0.40 &67.4 & 67.4 & 6.9E-4 & O(0.83) &  PISN  & 2.51 & 9.89 & 16.0 \\
230 & 0.1 $Z_\odot$ & 0.2 & 0.37 &69.4 & 69.4 & 7.1E-4 & O(0.83) &  PISN & 1.30 & 9.59  & 0.49 \\
240 & 0.1 $Z_\odot$ & 0.2 &0.37 & 71.0 & 71.0 & 7.3E-4 & O(0.83) &  PISN & 1.44 & 9.59  & 0.58 \\
250 & 0.1 $Z_\odot$ & 0.2 & 0.36 &72.6 & 72.6 & 7.5E-4 & O(0.83) &  PISN & 1.62 & 9.60  & 0.74 \\
300 & 0.1 $Z_\odot$ & 0.2 & 0.36 &75.7 & 75.7 & 7.8E-4 & O(0.83) &  PISN & 2.01 & 9.62  & 1.25 \\
350 & 0.1 $Z_\odot$ & 0.2 & 0.36 &89.9 & 89.9 & 9.2E-4 & O(0.83) &  PISN & 3.61 & 9.67  & 7.88 \\
400 & 0.1 $Z_\odot$ & 0.2 & 0.36 &108.2 & 108.2 & 1.1E-3 & O(0.83) &  PISN & 6.09 & 9.77  & 29.1 \\
410 & 0.1 $Z_\odot$ & 0.2 & 0.36 &114.2 & 114.2 & 1.2E-3 & O(0.83) &  PISN & 7.03 & 9.83  & 40.2 \\
420 & 0.1 $Z_\odot$ & 0.2 & 0.36 &115.5 & 115.5 & 1.1E-3 & O(0.83) &  CC & 0 & ---  & 0 \\
\hline
230 & 0.1 $Z_\odot$ & 0.3 & 0.50 &67.9 & 67.9 & 6.9E-4 & O(0.83) &  PPISN  & --- & --- &  0 \\
240 & 0.1 $Z_\odot$ & 0.3 & 0.49 &68.9 & 68.9 & 6.9E-4 & O(0.83) &  PISN  & 1.11 & 9.57 & 0.35 \\
250 & 0.1 $Z_\odot$ & 0.3 & 0.49 &71.4 & 71.4 & 7.6E-4 & O(0.83) &  PISN  & 0.40 & 9.54  & 0.053 \\
260 & 0.1 $Z_\odot$ & 0.3 & 0.48 &73.1 & 73.1 & 7.5E-4 & O(0.83) &  PISN  & 1.66 & 9.60  &  0.78 \\
280 & 0.1 $Z_\odot$ & 0.3 & 0.47 &76.2 & 76.2 & 7.8E-4 & O(0.83) &  PISN  & 7.43 & 9.79 & 1.22 \\
300 & 0.1 $Z_\odot$ & 0.3 & 0.47 &80.1 & 80.1 & 8.2E-4 & O(0.84) &  PISN  & 2.43 & 9.63 & 2.22 \\
350 & 0.1 $Z_\odot$ & 0.3 & 0.45 &85.0 & 85.0 & 8.7E-4 & O(0.84) &  PISN  & 3.01 & 9.65 &  4.54 \\
400 & 0.1 $Z_\odot$ & 0.3 & 0.42 &102.8 & 102.8 & 1.1E-3 & O(0.84) &  PISN  & 4.34  & 9.74 & 21.4 \\
430 & 0.1 $Z_\odot$ & 0.3 & 0.42 &113.8 & 113.8 & 1.2E-3 & O(0.83) &  PISN  & 6.95  & 9.84  &  39.4 \\
440 & 0.1 $Z_\odot$ & 0.3 & 0.42 &118.3 & 118.3 & 1.0E-3 & O(0.83) &  CC  & 0  & ---  & 0\\

		\hline \hline
	\end{tabular}
\end{table*}

\begin{figure}
    \centering        
    \includegraphics[width=1.0\linewidth]{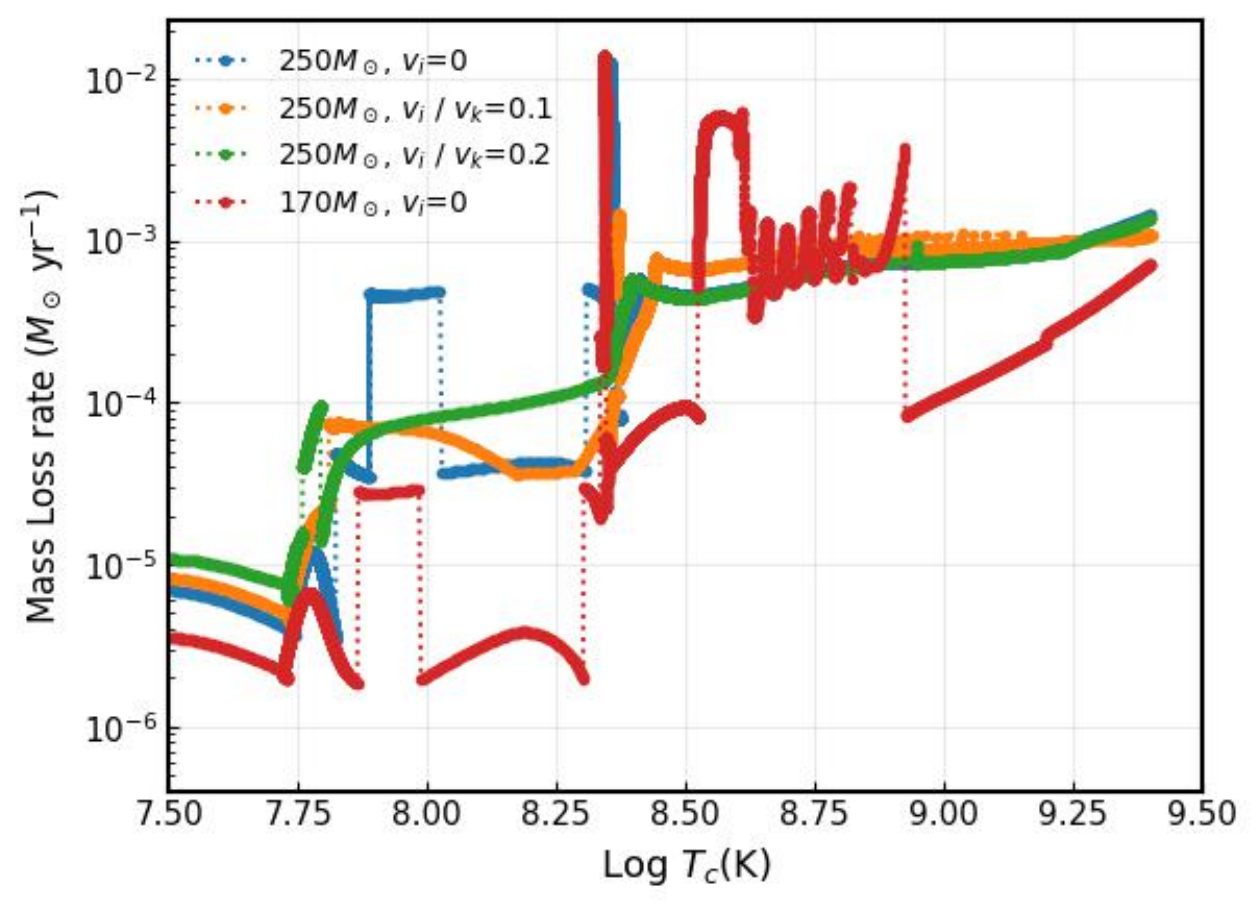}    
    \caption{Mass loss rates for the models in Figure \ref{fig:M_Tc}. }
    \label{fig:Mdot_Tc}
\end{figure}

\subsection{Mass loss histories}
\label{sec:masslossresults}

 Using some 0.1 $Z_\odot$ models, i.e., 170 $\msun$ $v_{\rm i}/v_{\rm k}= 0$ and
 250 $\msun$ $v_{\rm i}/v_{\rm k}= 0, 0.1$ and 0.2 models, we now describe 
 typical mass loss histories. All of these models end as PISNe
 though their final masses are different. Among them, the 170 $\msun$ model
 ends as a He-rich PISN while all the 250 $\msun$ models become
 O-rich PISNe. 
 
  Total mass evolution as a function of central temperature for these models is shown in
 Figure \ref{fig:M_Tc}. We find from this figure that mass loss is large
 at the end of H- and He-burning stages, which correspond to 
 $\log T_{\rm c} \sim $ 7.8 (K) and 8.3 (K), respectively.
 During the end of the H-burning stage, larger $v_{\rm i}/v_{\rm k}$ models
 tend to have larger mass loss rate. This is related to the surface composition 
 shown in Figure \ref{fig:X_Tc} and the size of core mass. Note that we define the CO core as the region inside the mass coordinate with $X(^{4}\rm He)>0.1$ inside the Helium layer.
 Figure \ref{fig:Mdot_Tc} shows the mass loss rates in these periods.
 
  With the effect of rotational matter mixing, the convective core is larger 
 for faster rotating models. For example, at  $\log T_{\rm c} \sim $ 7.8 (K),
 where the central hydrogen mass fraction is about 0.1, convective core masses
 for the $v_{\rm i}/v_{\rm k}= 0, 0.1$ and 0.2 models
 are about 140, 160 and 170$\msun$, respectively. Larger convective cores mean
 larger He-core masses and thus a star can be a He-star more easily with mass loss.
  As shown in Figure  \ref{fig:X_Tc},
 rotating models become He-rich at the surface in this stage, and the mass-loss rates
 become larger than that of the equivalent non-rotating models (Figure \ref{fig:Mdot_Tc}).
 In this particular example, the 250 $\msun$ $v_{\rm i}/v_{\rm k}= 0.2$ model
 becomes a Wolf-Rayet star in this stage.

 During the early He-burning stage, the evolutionary timescale is shorter than the
 mass loss timescale for this metallicity, and thus the total mass is roughly constant.
 However, at the end of the He-burning stage, stellar luminosity increases as the stellar
 core shrinks and mass loss rates increase again. Because of this enhanced mass loss,
 all the stars shown here become WR stars, though the detailed mass loss histories
 are rather complicated. For example, the 250 $\msun$ $v_{\rm i}/v_{\rm k}= 0$ model
 has a larger mass loss rate at the and of He-burning than rotating models,
 since the surface hydrogen mass fraction is larger. Because of this effect,
 the final masses of 250 $\msun$ $v_{\rm i}/v_{\rm k}= 0$ and 0.2 models end up being similar. The 250 $\msun$ $v_{\rm i}/v_{\rm k}= 0.1$ model has larger final
 mass, since it has smaller hydrogen mass fraction, but has larger He core mass
 than the non-rotating model.

  Since the 170$\msun$ model has lower luminosity than the 250$\msun$
  models, the mass loss rates are smaller in general and this model ends up as a He-rich
  star. In Figure \ref{fig:Mdot_Tc} we find rather large variation of the 
  mass loss rate for $\log T_{\rm c} = 8.5 \sim 8.8$. This is caused by the
  increase of the luminosity due to the shrinkage of stellar radius as the star
  becomes a WR star. Though we do not go into detail here, this variation
  may produce some observational signatures after the supernova explosion.

   Unlike the 0.1 $Z_\odot$ models, all the SMC and LMC models become O-rich stars
   in the PISN mass range as shown in Tables 2 and 3, because radiative mass loss
   is large enough to make these stars WR stars before the end of the H-burning 
   stage.

\subsection{Evolution of 
 Rotation and Angular momentum distribution}
\label{sec:rotation}

\begin{figure}
    \centering
    \includegraphics[width=1.0\linewidth]{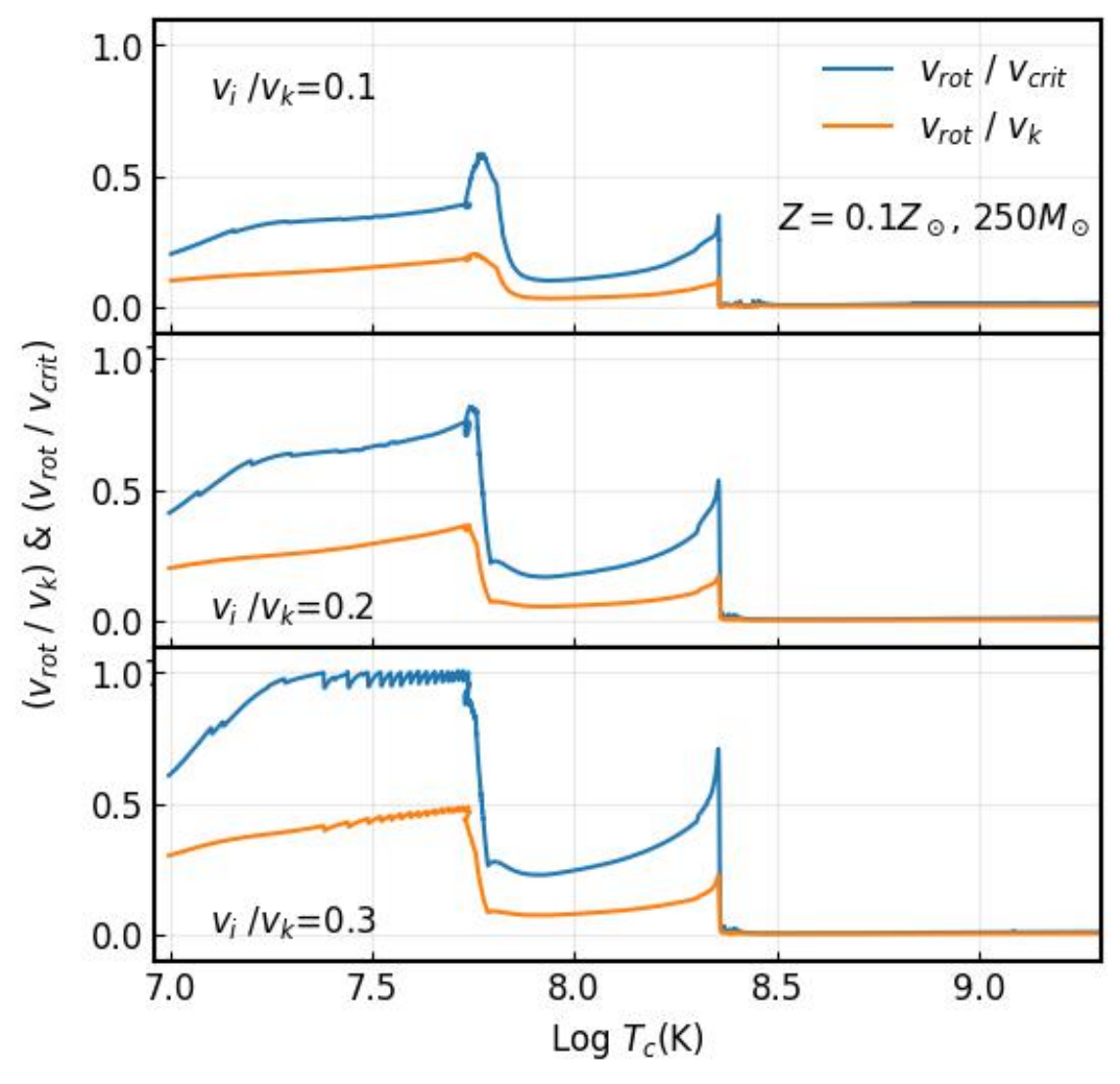}    
    \caption{$v_{\rm rot}/v_{\rm crit}$ and $v_{\rm rot}/v_{\rm k}$ 
    for 0.1$Z_\odot$, 250$\msun$ models with $v_{\rm i}/v_{\rm k} =0.1$, 0.2 and 0.3.}
    \label{fig:vrot_Tc}
\end{figure}

 As mentioned above, stellar rotation affects mass loss histories
 and the mass range for the PISN explosion. In this subsection we describe in
 more detail the evolution of surface rotation velocity and internal angular
 momentum distribution.

 One of the rotational effects is the enhancement of mass loss rates 
 according to equation (1). To see the effects of this, Figure
 \ref{fig:vrot_Tc} shows the evolution of $v_{\rm rot}/v_{\rm crit}$ as well as
 $v_{\rm rot}/v_{\rm k}$ for rotating 0.1 $Z_\odot$, 250 $\msun$ models.
 The $v_{\rm i}/v_{\rm k} $ = 0.1 and 0.2 models shown in this figure were
 discussed in the previous subsection. As we can see, $v_{\rm rot}/v_{\rm crit}$ 
 is always far below 1.0 for these models
 and the mass loss enhancement is insignificant.

 One may wonder, however, if we assume faster initial rotation, will $v_{\rm rot}$
 be larger. We show in the bottom panel just such a case for 
 $v_{\rm i}/v_{\rm k} $ = 0.3. Indeed $v_{\rm rot}/v_{\rm crit}$ reaches 1.0
 before the main sequence stage. However, after some period of the enhanced 
 mass loss stage, the star loses angular momentum and $v_{\rm rot}/v_{\rm crit}$ 
 never reaches 1.0 again after the main-sequence stages. 
 The mass loss rate of this model is persistently larger than slower
 rotating ones, and the final mass roughly corresponds to the lowest mass
 PISN, which ejects only a small amount of $^{56}$Ni.

\begin{figure}
    \centering
    \includegraphics[width=1.0\linewidth]{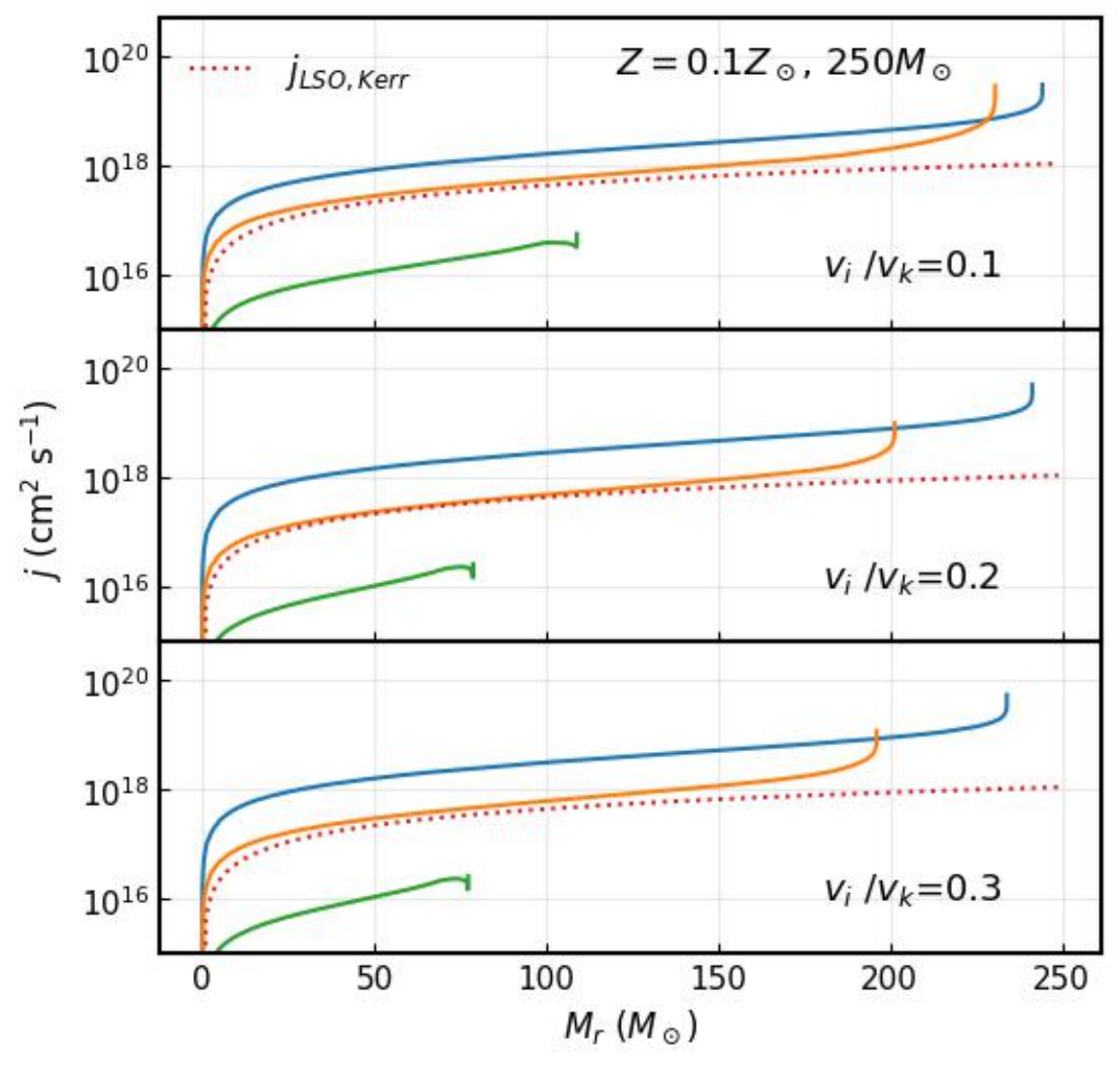}      
    \caption{Specific angular momentum $j$ for the 0.1$Z_\odot$, 250 $M_\odot$ models
    with $v_{\rm i}/ v_{\rm k}=0.1$, 0.2 and 0.3.}
    \label{fig:j_Mr}
\end{figure}

 In Figure \ref{fig:j_Mr}, we show the distribution and time evolution of
 specific angular momentum $j$ for the same models as in Figure \ref{fig:vrot_Tc}.
 Three solid lines correspond to $j$ at the epochs when the central
 hydrogen mass fraction becomes $X=0.5$, $X=0.1$ and the central helium mass fraction becomes
 $Y=0.01$. These three epochs corresponds to the middle and end of the hydrogen
 burning stage, and the end of the helium burning stage, respectively.
  Each panel in this figure also show a dashed curve labeled $j_{\rm LSO, Kerr}$. 
 This is the specific angular momentum needed to get into the last stable orbit around a 
 maximally rotating Kerr-black hole of rest-mass equal to the mass coordinate
 \citep{Bardeen1972ApJ...178..347B}. This curve gives a rough measure 
 of the amount of angular momentum necessary to be a GRB progenitor,
 and we may call the model fast rotating if $j$ of the model exceeds $j_{\rm LSO, Kerr}$
  \citep{Yoon2005A&A...443..643Y,Yoon2006A&A...460..199Y}. 
 This figure show that the angular momentum loss during and after the helium burning
 stages is efficient and the star becomes a slow rotator even for the  
 $v_{\rm i}/v_{\rm k} $ = 0.3 model. In fact we find that all the PISNe models
 shown in this paper are slow rotators, which means that CHE does not occur 
 for this mass and metallicity range. We note that these results depend on the
 mass loss rate for hydrogen poor stars, which is rather uncertain due to
 the paucity of observations.

\subsection{
Internal Abundance Evolution Before PISN Explosion}
\label{sec:internal}

\begin{figure}
    \centering
    \includegraphics[width=1.0\linewidth]{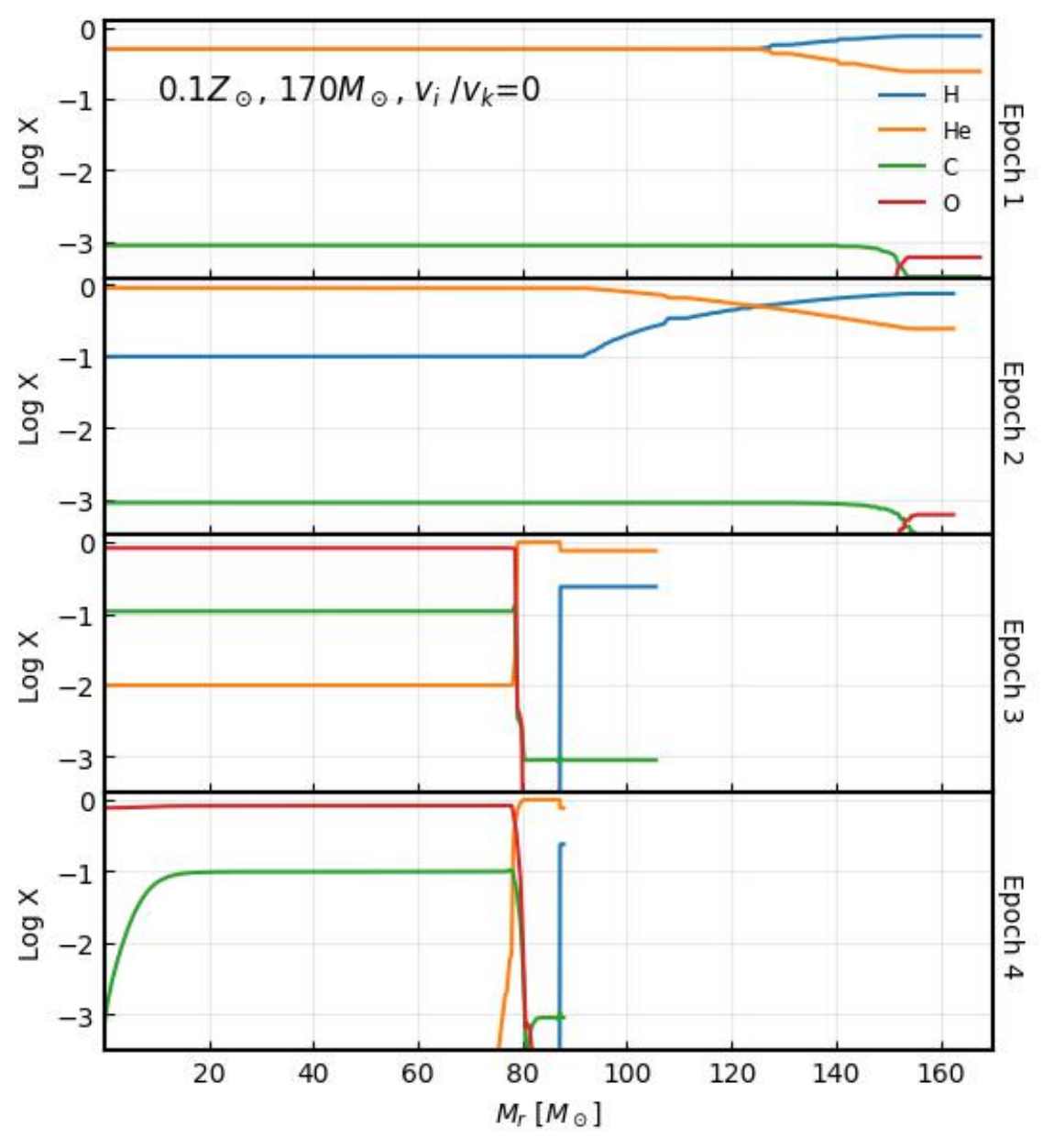}     
    \caption{Abundance evolution for the $Z=0.1Z_\odot$, 170 $M_\odot$, $v_{\rm i}/ v_{\rm k}=0$ 
    model for the 4 epocs defined in the text.
    }
    \label{fig:X_Mr_1}
\end{figure}

 In this subsection, we explain the evolution of internal abundance distributions
 using the same models shown in Section \ref{sec:masslossresults}, namely,
 $Z=0.1Z_\odot$, 170$\msun$ ($v_{\rm i}=0$) and 
 250 $\msun$ ($v_{\rm i}/ v_{\rm k}=0, 0.1, 0.2$) models. 
 Figures 7 to 10 show abundance distributions for these models
 for 4 epochs which are defined by the time when the central hydrogen mass fractions
 are $X=0.5$ and 0.1 (Epochs 1 and 2, respectively), central helium mass
 fraction is $Y=0.01$ (Epoch 3), and at the central temperature is 
 $\log T_{\rm c} = 9.2 $ (K) (Epoch 4)
 which corresponds to the beginning of pair-instability collapse. 

 First we compare two non-rotating models in Figures 7 and 8.
 Until Epoch 2, these models have similar abundance patterns, though
 the 250 $\msun$ model has a larger convective core mass as we can see from the
 flat distributions of H and He in the core. The evolution after Epoch 3 
 appears different. This is because the 250 $\msun$ model has a larger mass-loss 
 rate because of larger luminosity. As a result, the star becomes
 a WR star during the middle of the helium burning stage. Then, the mass loss
 rate becomes even larger, which can be seen in Figure 4, 
 and the model loses the entire He-rich envelope before the start of Epoch 3. 
 The opposite applies to 
 the 170 $\msun$ model and this model maintains a thick He envelope until Epoch 4.

 Next we describe rotation effects using Figures 8 to 10 for 
 250 $\msun$ models with different rotation velocities.
 From these figures we find that faster rotating models have larger
 convective core masses because of rotational matter mixing.
 As a result, faster rotating models tend to have larger mass-loss rates,
 and they become WR stars earlier, which we can verify in the figures for Epoch 2.
 We note that as mentioned in Section \ref{sec:masslossresults}, this 
 mass-loss enhancement is not caused by the rotation induced mass loss, but simply
 by the increase in luminosity.
 
 The slowly rotating model with $v_{\rm i}/ v_{\rm k}=0.1$ in Figure 9,
 has a larger final core mass than the non-rotating model in Figure 8
 because the increase of core mass by rotational mixing is more
 effective than the increase of mass-loss rate. On the other hand, 
 for faster rotating models with $v_{\rm i}/ v_{\rm k} \gtrsim 0.2$,
 the effect of mass loss is more important and the final mass becomes
 smaller for faster rotation.  

 Though not shown in figures, with our choice of mass loss rates, 
 low mass PISN models have He-rich envelopes only for low metallicity:
 $Z \lesssim 0.1 Z_\odot$. As shown in Table 3, all our PISNe models
 with  $Z \geq 0.2 Z_\odot$ have O-rich envelopes, even for the lowest 
 mass models because of larger radiative mass loss at higher metallicities.

\begin{figure}
    \centering
    \includegraphics[width=1.0\linewidth]{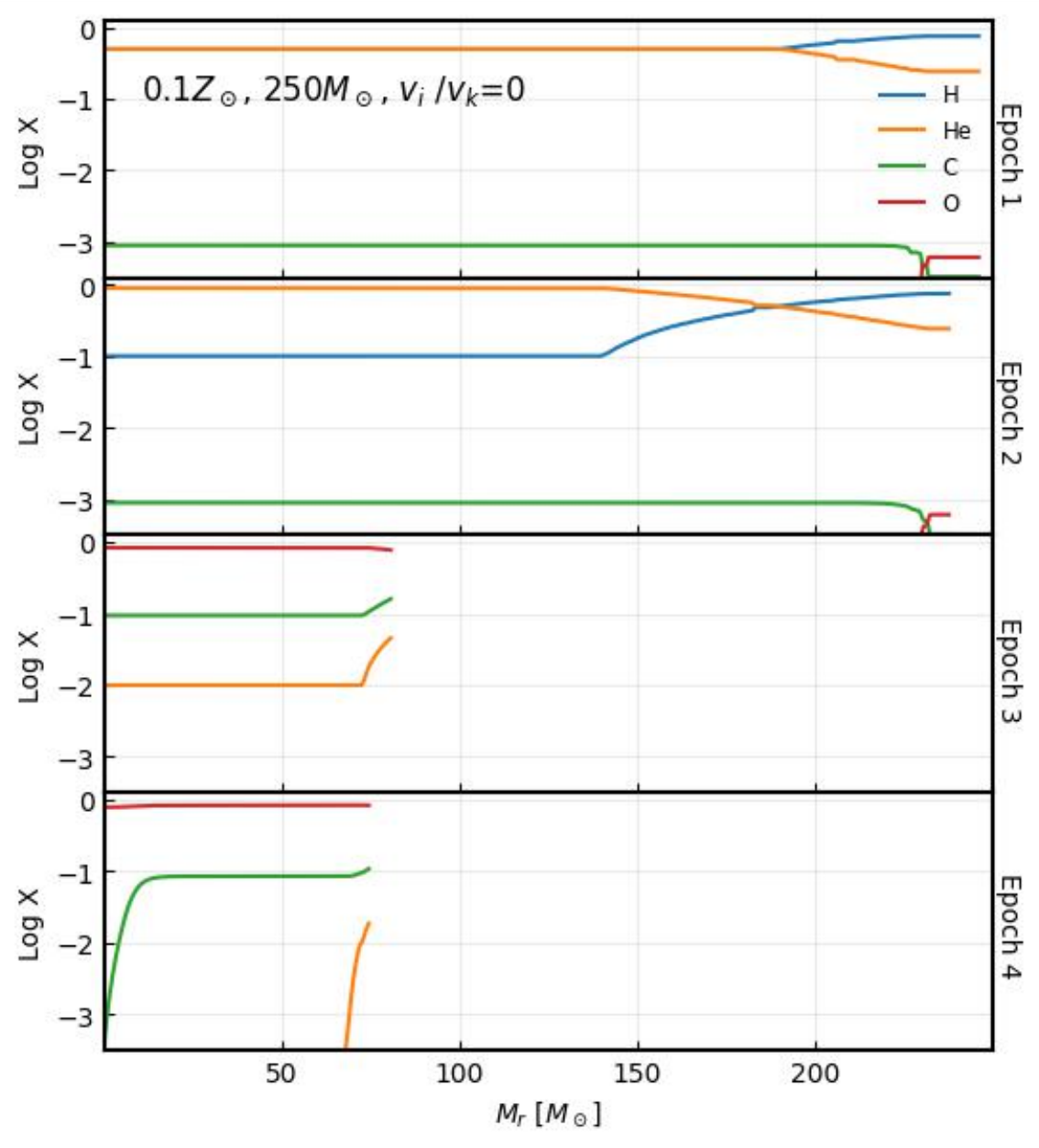}    
    \caption{Same as Figure \ref{fig:X_Mr_1} but for the 
    $Z=0.1Z_\odot$, 250 $M_\odot$, $v_{\rm i}/ v_{\rm k}=0$ model.
    }
    \label{fig:X_Mr_2}
\end{figure}

\begin{figure}
    \centering
    \includegraphics[width=1.0\linewidth]{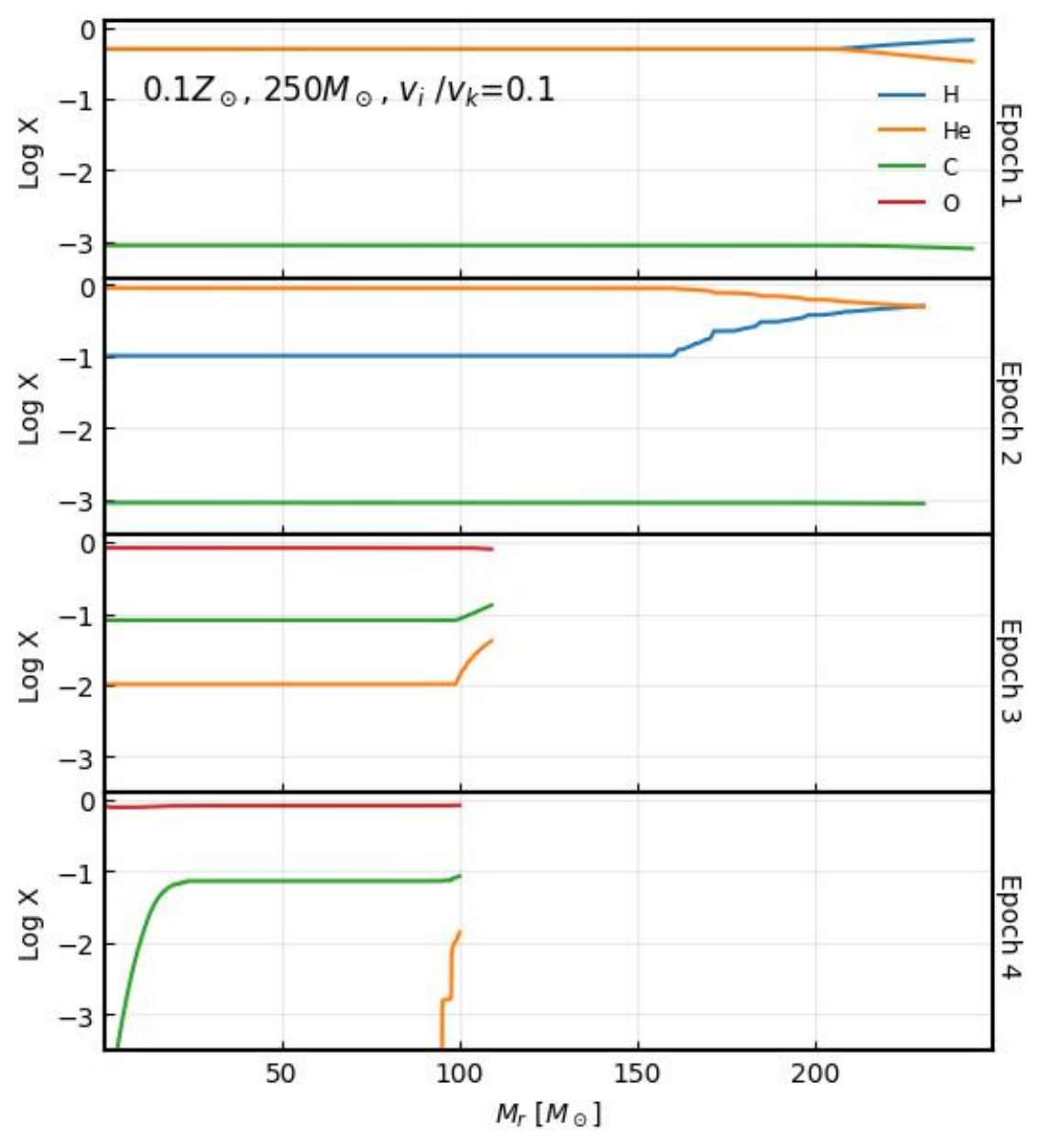}    
    \caption{Same as Figure \ref{fig:X_Mr_1} but for the 
    $Z=0.1Z_\odot$, 250 $M_\odot$, $v_{\rm i}/ v_{\rm k}=0.1$ model.
    }
    \label{fig:X_Mr_3}
\end{figure}

\begin{figure}
    \centering
    \includegraphics[width=1.0\linewidth]{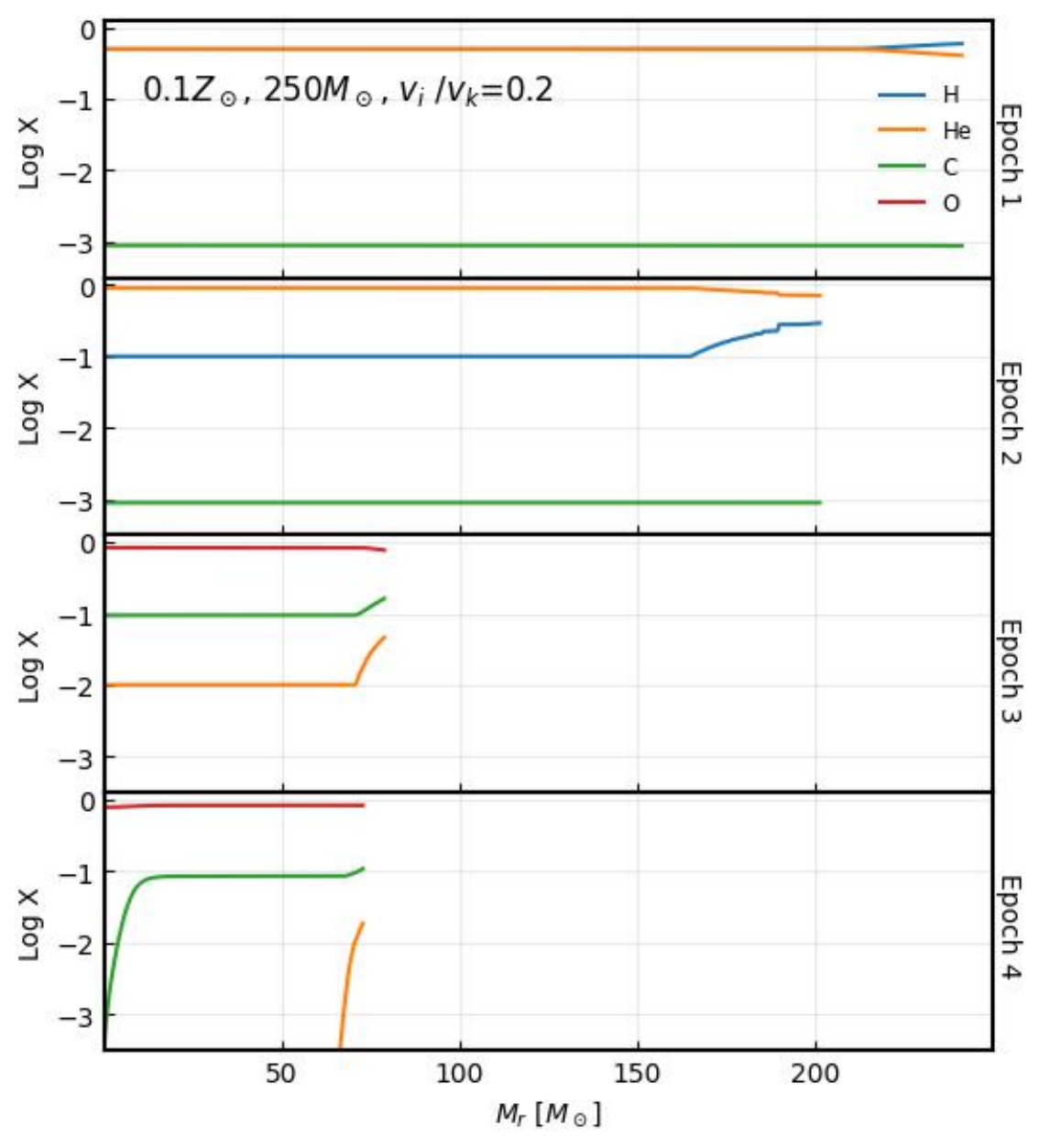}    
    \caption{Same as Figure \ref{fig:X_Mr_1} but for the 
    $Z=0.1Z_\odot$, 250 $M_\odot$, $v_{\rm i}/ v_{\rm k}=0.2$ model.
    }
    \label{fig:X_Mr_4}
\end{figure}

\begin{figure*}
    \centering

    \includegraphics[width=0.49\linewidth]{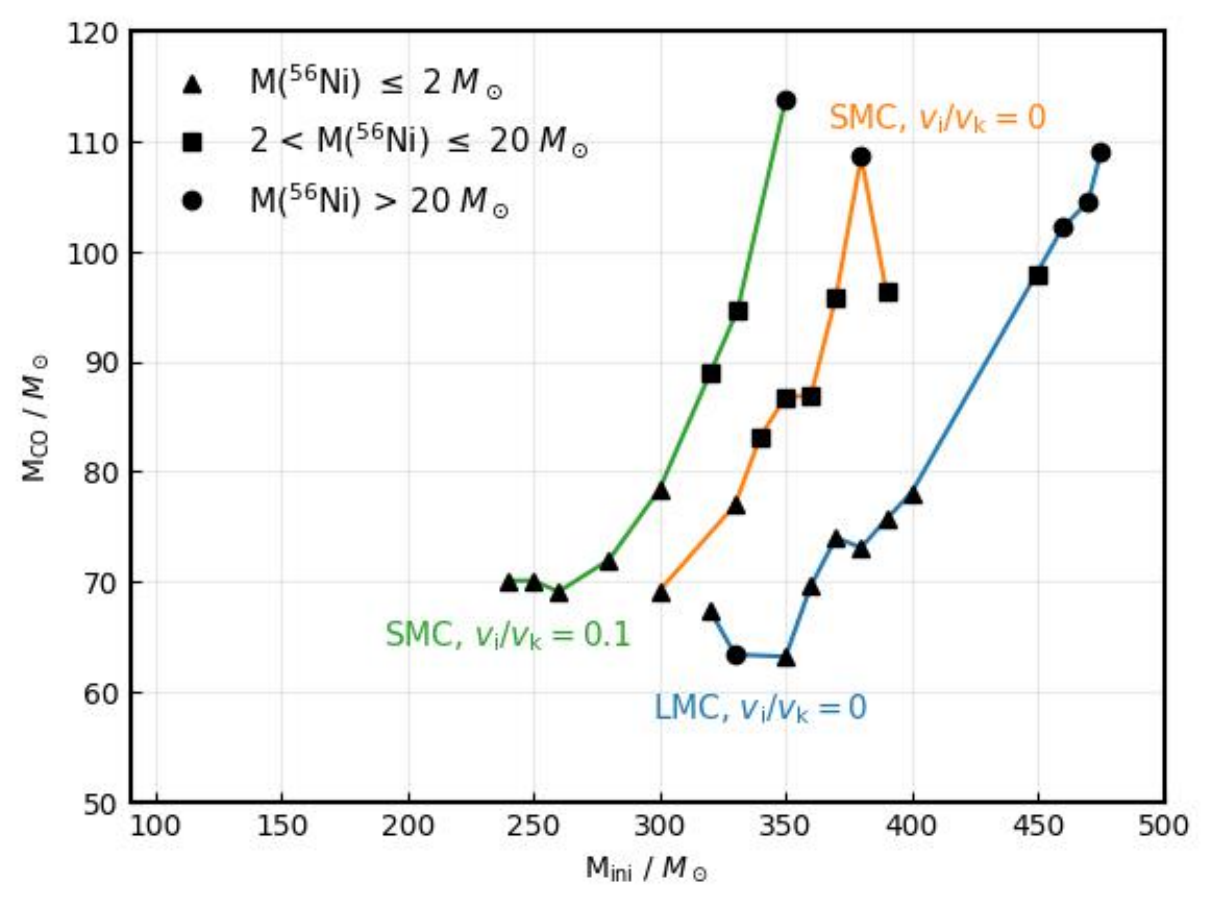}    
    \includegraphics[width=0.49\linewidth]{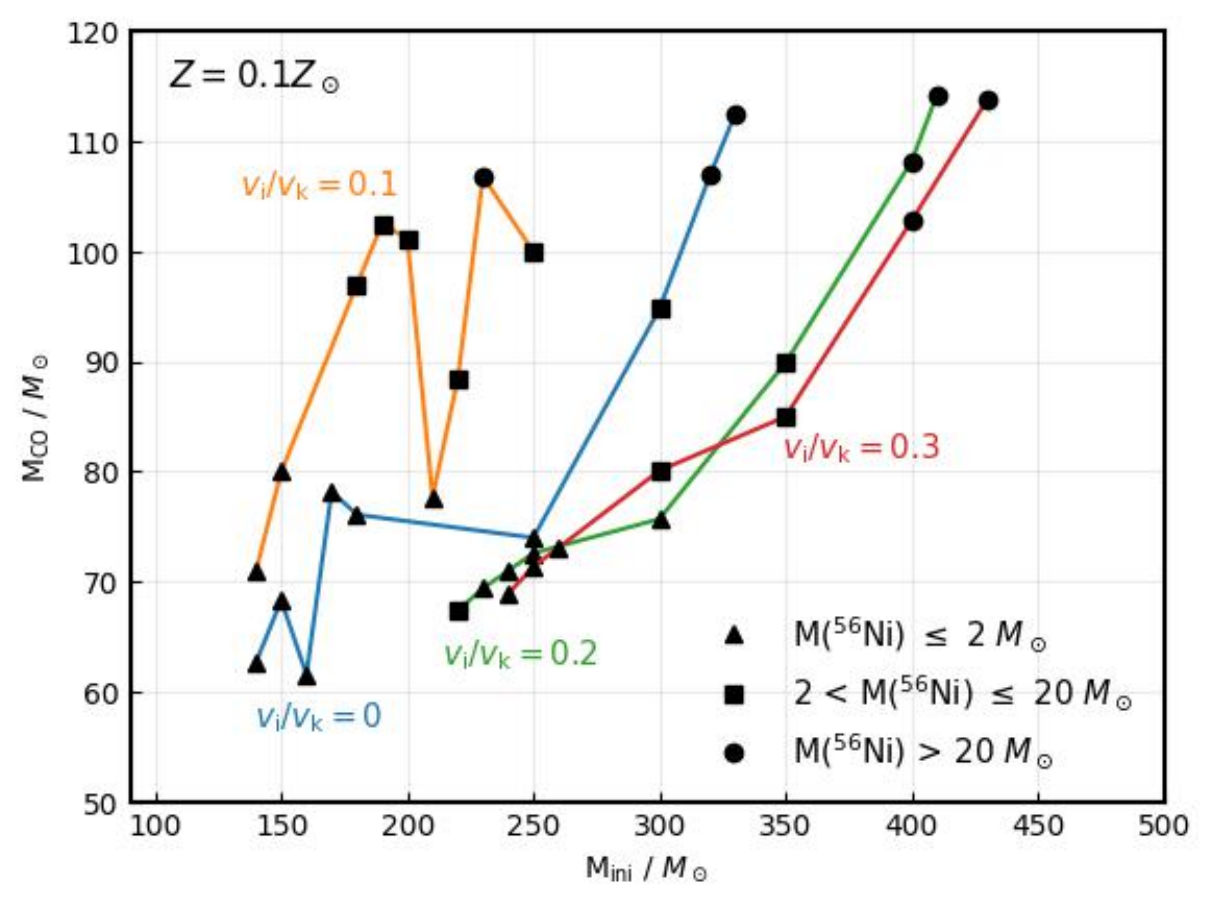}   
    \caption{Summary figure showing the CO core mass as a function of the initial mass for the exploding models. The symbols show the amount of $^{56}$Ni produced in the explosion, while the colored lines connect models with the same parameters besides mass (subsections of Tables \ref{tab:hydro}, \ref{tab:hydro1}). The left panel shows SMC and LMC models, while the right panel shows Z$=0.1 \zsun$ models.
    }
    \label{fig:MCOs}
\end{figure*}

\subsection{Hydrodynamical calculations and PISN explosions}
\label{sec:hydroresults}

For most models, when the central temperature reaches $10^{9.2}$ K, we switch to the hydrodynamics code. The exceptions to this are the $v_i/v_k=0.2$. After this switch, all models collapse immediately, reaching their peak temperatures within 200 to 700 seconds. After this, the velocity reverses, and a shock forms, which propagates from the center of the star to the surface in 10-100 seconds. Maximum temperatures and explosion energies (Table \ref{tab:hydro}) fall in roughly the same range as in the Pop III case \citep[e.g.][]{takahashi2018}. Note that models which fail to explode because they pulse or collapse to a black hole are denoted by "PPISN" and "Collapse", respectively, in the final three columns of Table \ref{tab:hydro}. One difference to the Pop III case is that the $^{56}$Ni mass is higher for comparable CO core masses. This may derive from seed metals not present in the Pop III case (Subsection \ref{sec:nuc}), although it is hard to say for certain (Section \ref{sec:discussion}).

\subsection{Nucleosynthesis during PISN explosions}
\label{sec:nuc}

\begin{figure*}
    \centering

    \includegraphics[width=1.0\linewidth]{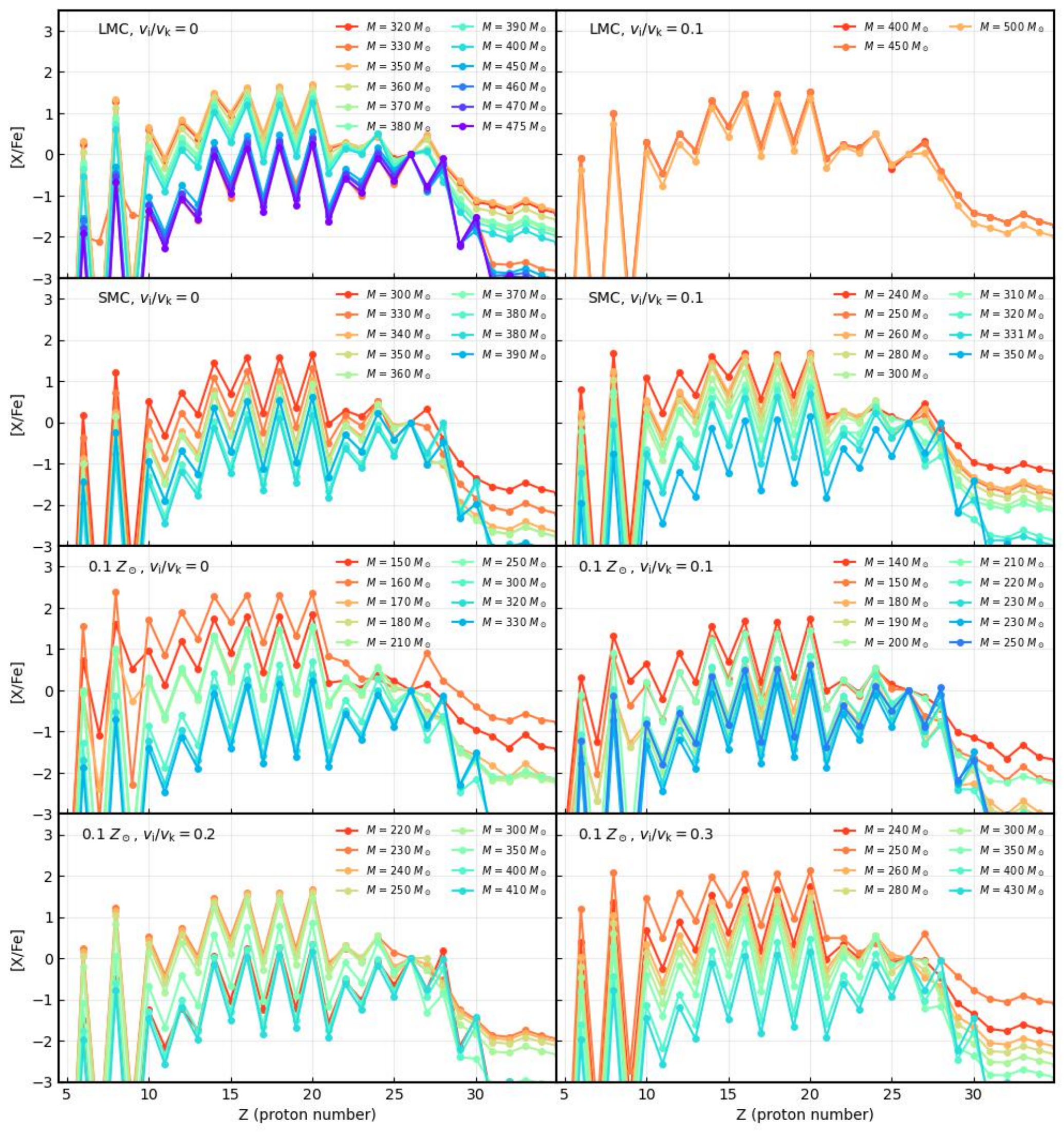}    
    \caption{Log elemental abundances relative to Iron, relative to solar \citep{asplund2009} for the explosions in Tables \ref{tab:hydro}, \ref{tab:hydro1}, where each panel corresponds to a single subsection of these two tables. The abundances are the results of the post processing, including decaying the ejecta for $10^{16}$ s (Subsection \ref{sec:hydromethods}).
    }
    \label{fig:yields}
\end{figure*}

After performing the hydrodynamical simulations with 153 isotopes, we then post process the trajectories of those simulations with a larger network (300 isotopes). As the temperature increases, isotopes above the iron group begin to photodissintegrate. Note that this process does not occur in the Pop III case as there are no heavy isotopes. By Log T $\approx$ 9.5, the composition has moved away from stability towards the proton rich side (p side), forming a continuous distribution in the neutron-proton plane. As the temperature increases further, the composition shifts to higher mass and towards the neutron rich side, so that, at maximum temperature (Log T $\approx$ 9.8, in the extreme case), the composition spans nearly the entire 300 isotope network. Once the temperature and density decrease, the distribution retreats to the p side, before eventually decaying back to stability. Figure \ref{fig:MCOs} shows the Nickel mass (symbols) of all models in the plane of the CO core mass and the initial mass. The Nickel mass is generally thought to depend only on the CO core mass, but as we discuss below (subsection \ref{sec:peculiar}), using a finer grid and including metal enriched and rotating progenitors has revealed more complex behavior. Figure \ref{fig:yields} shows the elemental yields for all models (grouped similarly to in Table \ref{tab:hydro}) and Table \ref{tab:yields} shows isotopic yields for the M $=475$ $\msun$, Z = LMC, $v_i=0$ model. As in the Pop III case, the main signature of the PISNe is the sawtooth pattern derived from the preference for even elements \citep{Heger2002ApJ...567..532H,Umeda2002ApJ...565..385U,takahashi2018}. This sawtooth pattern extends to heavier elements for more massive cores (Figure \ref{fig:yields}) which reach higher peak temperatures (Table \ref{tab:hydro}).

\begin{table*}[t]
	\centering
	\caption{Yields (Y [$\msun$]) of all stable isotopes up to $^{79}$Be for the 300 isotope post processing of M $=475$ $\msun$, Z = LMC, $v_i=0$. The upper section contains the yields after the composition has fully decayed. The lower section contains the yields for radioactive isotopes with Y$>10^{-6}$ at the end of the hydrodynamics calculation ($t\approx 8000$ s). Yield tables for all models are available in the supplementary material.}
	\label{tab:yields}
	\begin{tabular}{cl|cl|cl|cl|cl|clcccccccccccccccccccc} 
		    \hline \hline
    		I & Y & I & Y & I & Y & I & Y & I & Y & I & Y    \\
    		\hline
$^1$H & 1.568E-10 & $^{18}$O & 1.124E-8 & $^{33}$S & 1.037E-2 & $^{46}$Ca & 2.991E-8 & $^{54}$Fe & 4.659E-1 & $^{68}$Zn & 4.226E-6\\
$^2$H & 6.128E-19 & $^{19}$F & 8.658E-9 & $^{34}$S & 5.826E-2 & $^{48}$Ca & 6.058E-7 & $^{56}$Fe & 32.44 & $^{70}$Zn & 9.193E-8\\
$^3$He & 6.061E-19 & $^{20}$Ne & 1.489 & $^{36}$S & 2.509E-5 & $^{45}$Sc & 2.347E-5 & $^{57}$Fe & 5.743E-1 & $^{69}$Ga & 7.638E-7\\
$^4$He & 8.161E-1 & $^{21}$Ne & 5.563E-4 & $^{35}$Cl & 3.331E-3 & $^{46}$Ti & 5.224E-4 & $^{58}$Fe & 8.102E-3 & $^{71}$Ga & 3.031E-7\\
$^6$Li & 6.338E-21 & $^{22}$Ne & 1.448E-5 & $^{37}$Cl & 1.544E-3 & $^{47}$Ti & 1.594E-5 & $^{59}$Co & 1.712E-2 & $^{70}$Ge & 4.181E-6\\
$^7$Li & 9.956E-24 & $^{23}$Na & 4.346E-3 & $^{36}$Ar & 2.656 & $^{48}$Ti & 1.900E-2 & $^{58}$Ni & 1.085 & $^{72}$Ge & 8.715E-7\\
$^9$Be & 2.195E-27 & $^{24}$Mg & 1.39 & $^{38}$Ar & 3.670E-2 & $^{49}$Ti & 6.719E-4 & $^{60}$Ni & 3.153E-1 & $^{73}$Ge & 9.598E-8\\
$^{10}$B & 3.289E-20 & $^{25}$Mg & 2.919E-2 & $^{40}$Ar & 2.011E-5 & $^{50}$Ti & 1.083E-6 & $^{61}$Ni & 1.186E-2 & $^{74}$Ge & 6.836E-7\\
$^{11}$B & 6.170E-15 & $^{26}$Mg & 5.451E-2 & $^{39}$K & 4.515E-3 & $^{50}$V & 8.274E-7 & $^{62}$Ni & 1.129E-1 & $^{75}$As & 2.201E-7\\
$^{12}$C & 7.847E-1 & $^{27}$Al & 4.020E-2 & $^{40}$K & 5.176E-6 & $^{51}$V & 1.085E-3 & $^{64}$Ni & 1.205E-5 & $^{74}$Se & 5.253E-8\\
$^{13}$C & 6.187E-8 & $^{28}$Si & 16.69 & $^{41}$K & 4.822E-4 & $^{50}$Cr & 3.284E-3 & $^{63}$Cu & 7.591E-5 & $^{76}$Se & 2.298E-7\\
$^{14}$N & 6.086E-7 & $^{29}$Si & 3.448E-2 & $^{40}$Ca & 2.895 & $^{52}$Cr & 3.497E-1 & $^{65}$Cu & 6.949E-5 & $^{77}$Se & 4.503E-8\\
$^{15}$N & 3.161E-8 & $^{30}$Si & 5.274E-2 & $^{42}$Ca & 1.130E-3 & $^{53}$Cr & 1.776E-2 & $^{64}$Zn & 3.959E-4 & $^{78}$Se & 3.905E-7\\
$^{16}$O & 33.74 & $^{31}$P & 1.800E-2 & $^{43}$Ca & 3.755E-5 & $^{54}$Cr & 9.630E-5 & $^{66}$Zn & 1.175E-3 & $^{79}$Br & 1.067E-7\\
$^{17}$O & 5.795E-6 & $^{32}$S & 12.69 & $^{44}$Ca & 2.165E-3 & $^{55}$Mn & 7.559E-2 & $^{67}$Zn & 1.616E-6 & &\\
\hline
$^{22}$Na & 9.492E-6 & $^{41}$Ca & 4.186E-4 & $^{51}$Cr & 1.079E-3 & $^{60}$Fe & 4.556E-6 & $^{57}$Ni & 5.301E-1 & $^{65}$Zn & 5.565E-5\\
$^{26}$Al & 2.218E-3 & $^{43}$Sc & 2.432E-6 & $^{52}$Mn & 1.221E-1 & $^{55}$Co & 5.733E-2 & $^{59}$Ni & 1.518E-2 & $^{66}$Ga & 6.670E-4\\
$^{32}$P & 8.295E-6 & $^{44}$Ti & 2.154E-3 & $^{53}$Mn & 1.775E-2 & $^{56}$Co & 3.283E-3 & $^{63}$Ni & 5.950E-6 & $^{66}$Ge & 6.133E-5\\
$^{33}$P & 6.040E-6 & $^{45}$Ti & 3.252E-6 & $^{54}$Mn & 3.596E-6 & $^{57}$Co & 4.413E-2 & $^{61}$Cu & 3.568E-3 & &\\
$^{35}$S & 3.848E-6 & $^{48}$V & 1.022E-4 & $^{52}$Fe & 2.267E-1 & $^{58}$Co & 2.437E-5 & $^{62}$Cu & 1.774E-4 & &\\
$^{36}$Cl & 1.235E-5 & $^{49}$V & 6.702E-4 & $^{55}$Fe & 1.793E-2 & $^{60}$Co & 2.664E-5 & $^{64}$Cu & 3.005E-6 & &\\
$^{37}$Ar & 1.283E-3 & $^{48}$Cr & 1.888E-2 & $^{59}$Fe & 5.221E-5 & $^{56}$Ni & 32.44 & $^{62}$Zn & 1.057E-1 & &\\
		\hline \hline
	\end{tabular}
\end{table*}

\subsection{Peculiar phenomena}
\label{sec:peculiar}

 In this subsection, we summarize some peculiar phenomena we have found during 
 the calculations. It is usually expected that there is a concrete mass region for PISNe
 for given metallicity and initial rotational velocity, and larger initial mass
 leads to larger CO core mass and larger explosion energy (and $^{56}$ Ni mass)
 in the range. However, we find that this is not always the case because mass loss
 histories can be complicated. Especially near the border of the PISN range, there is 
 often no simple relation between the initial mass and the final outcome.
 For example, in $0.1 Z_\odot$ $v_{\rm i}=0$ models, the relation between the 
 initial mass and explosion energy is not monotonic below $170M_\odot$. 
 As a result, the $160 \msun$ model becomes a PPISN while the $150 \msun$ model
 becomes a PISN. We find similar "inversion" near the upper mass limits
 in the SMC and $0.1 Z_\odot, v_{\rm i}/ v_{\rm k}=0.1$ models.
 Related to this, we have core collapse for 330 and 340 $\msun$ but a PISN explosion
 for 350 $\msun$ in the SMC, $v_{\rm i}/ v_{\rm k}=0.1$ models. 
 Because of the abnormality of these findings, we also calculated a 331 $\msun$ model
 which exploded (Table 2). These results suggest that it is not 
 a simple task to find an exact border for the PISN range when the effects of mass-loss are large.

 We also find that some models, such as SMC, 330$\msun$, $v_{\rm i}/ v_{\rm k}=0.1$,
 0.1 $Z_\odot$, 200 $\msun$, $v_{\rm i}=0$, and 0.1 $Z_\odot$, 
 170 $\msun$, $v_{\rm i}/ v_{\rm k}=0$, have core masses which are expected to result in PISNe, but they do not explode. These models emerge from the helium burning phase with low entropy, and thus they avoid the PISN region, despite their massive cores. We have verified the stability of these models in the hydrodynamical code and expect them to eventually undergo core-collapse and form black holes with masses right in the middle of the PISN mass gap.

 Finally, we mention another peculiar result for LMC, 330 $\msun$, $v_{\rm i}=0$
 (referred to as the 330 $\msun$ model below, for brevity).
 This model is at the lower edge of the PISNe range, and indeed the LMC, 340 $\msun$, $v_{\rm i}=0$
 becomes a PPISN. For the PPISN model, the first $T_{\rm c}$ peak above 
 $\log T_{\rm c}= 9.2 $ (K) is not high enough to blow up the entire star, and only 
 the outer envelope is ejected during the first "pulse". In the 330 $\msun$ model, on the other hand,
 the peak temperature is very high ($\log T_{\rm c}= 9.79 $ (K) )
 but slightly below the photo-dissociation temperature for Fe, and thus the star
 explodes with very high energy and large $^{56}$Ni production even though the
 CO-core mass is rather small ($M_{\rm CO} = 63.4 \msun$). We note that similar
 evolution is seen in the 0.1 $Z_\odot$, 220 $\msun, v_{\rm i}/ v_{\rm k}=0.1$ model,
 though the explosion energy is not as large as the 330 $\msun$ model. We have verified the properties of this explosion with much stricter numerical settings and we believe that it is real. We emphasize, however, that it likely occurs only in a very narrow mass range at the lower bound of the PISNe region, and should thus be a rare occurrence. 
 
 ~\vspace{7mm}
\subsection{Maximum $^{56}$ Ni mass}

 The maximum $^{56}$ Ni mass produced by PISNe is interesting, because it 
determines the maximum brightness for observed PISNe. Among the models we show in
Tables 2 and 3, the maximum $^{56}$Ni mass is 40.2 $\msun$ for the 
0.1 $Z_\odot$, 410 $\msun, v_{\rm i}/ v_{\rm k}=0.2$ model. This maximum mass
is roughly the same as previous Pop III PISNe results \citep{Heger2002ApJ...567..532H,takahashi2018},
which is reasonable because the nuclear burning during pair-instabillity collapse
should be roughly independent of metallicity.

 Since we show our results only with 10 or 5 $\msun$ intervals with respect to the
 initial mass, one may wonder if much higher $^{56}$Ni mass could be obtained if we calculated
 models with finer grids. We thus tried to calculate with
 much finer grids for several cases. However, we did not find any models with $^{56}$Ni
 higher than 40 $\msun$, this is partly because we do not necessary have larger 
 CO-core mass for a larger initial mass, as mentioned in the previous sub-section.

 In order to obtain continuous $M_{\rm CO}$ size samples, we made the following 
artificial  modification using the 0.1 $Z_\odot$, 230 $\msun, v_{\rm i}/ v_{\rm k}=0.1$ model.
The original model has $M_{\rm CO}=106.8$ and $^{56}$Ni = 25.2 $\msun$.
In order to have a larger final core, we artificially reduce the mass loss rate
by a constant factor only for the period when $\log T_{\rm c}= 8.5\sim 8.9$, so that we 
could obtain a homogeneous progenitor set with different $M_{\rm CO}$. 
The maximum possible $M_{\rm CO}$ to be a PISN obtained in this way is 118.3 $\msun$ 
and the results are shown in the Table 3 labeled as '$M_{\rm i} = $ 230B'.
As shown there, the maximum energy is $7.00 \times 10^{52}$ erg with $^{56}$Ni = 37.5 $\msun$. This exercise demonstrates that a $^{56}$ Ni mass upper limit of $\simeq $ 40 $\msun$ is likely quite general. 

\section{Discussion}
\label{sec:discussion}

 First we compare our results with Pop III PISN calculations in \citet{takahashi2018}.
 For the Pop III models, PISNe occurred in the CO core mass rage,
 $M_{\rm CO} \sim 72 - 134 \msun$, though the range depends on models. 
 From Table \ref{tab:hydro}, we find the range is lower for our metal-rich models,
 $M_{\rm CO} \sim 63 - 120 \msun$, though in some parameter sets, the lower bound is higher, $M_{\rm CO} \sim 71 \msun$ and the upper bound is lower, $M_{\rm CO} \sim 110 \msun$.
 
 There is also a difference in the relation between
 $M_{\rm CO}$ and $^{56}$Ni mass or explosion energy. Our metal-enriched PISNe models
 tend to have higher $^{56}$Ni masses and explosion energies than the Pop III models.
 Without further investigations, we cannot say, at this moment, that these differences are
 metallicity effects, since some adopted parameters including the 
 $^{12}C(\alpha,\gamma)^{16}O$ rate are different from \citet{takahashi2018}.
 Nevertheless, we can say that it is dangerous to simply assume that a metal-rich PISNe 
 is the same as a Pop III PISN with the same $M_{\rm CO}$.

 Comparisons of the PISN range with other previous models were made in Section 3.1 and
 we do not repeat it here. We simply note that the results depend on mass-loss rates, 
 which are rather uncertain, and in particular the
 WR mass-loss rates are very important.

 One of the significant differences in our results
 with previous work is that most our models become
 O-rich PISNe. We find He-rich PISNe only for metal poor ($Z \lesssim 0.1 Z_\odot$),
 low mass PISNe. If this is true, it would be difficult to find He-rich PISNe
 in nearby samples. This could be consistent with past observations that have not
 found any evidences for He-rich PISNe, and one promising PISN candidate, SN2018ibb, described
 in the next subsection appears to be O-rich. We even suggest that we might have already
 found such O-rich SNe already in the samples of many Type-I super luminous SNe
 (SLSNe). It is possible that
 we simply have overlooked such a possibility since O-rich PISNe models have not yet been reported.

 \subsection{Type Ic PISN candidate SN2018ibb}
 \label{sec:SN2018ibb}

 Very recently \citet{Schulze2023arXiv230505796S} reported that SN2018ibb is
 the best PISN candidate to date. Though they only say that it is Hydrogen poor,
 it is likely that it is Type Ic since they show evidence that this SN
 interacts with Oxygen-rich CSM. 

 If this was a He-rich SN Ib, it is difficult to explain by our models, 
 because SN2018ibb is bright and the expected $^{56}$Ni mass is about
 30 $\msun$. On the other hand, if this was a O-rich SN Ic, several models
 may fit, such as LMC $\sim 470 \msun$, SMC $\sim 380 \msun$, 0.1 $Z_\odot$ $\sim 300 \msun$
 for non-rotating models, and SMC $\sim 350 \msun$ and 0.1 $Z_\odot$ $\sim 230 \msun$
 for $v_{\rm i}/ v_{\rm k}=0.1$ models.

 For SN2018ibb, evidence of CSM interaction is observed at around 600 days after the 
 peak of bolometric luminosity. In the next subsection we discuss the possibilities
 of such an interaction in our model, though more detailed discussion about SN2018ibb will
 be given in a forthcoming paper (Nagele et al, in prep.).

\subsection{Late time mass loss and SN-CSM interaction}

 It is well known that some SNe show evidence of interaction with CSM 
 which has been ejected from the progenitors through mass loss. 
 For a usual WR star with masses below $30 \msun$, the mass loss rates
 are about $10^{-4} \msun$/ yr or below \citep[see e.g.][]{Yoon2017MNRAS.470.3970Y}, which is low enough to expect SN-CSM interaction to not be observable. 
 For SNe Ibn, the expected mass loss rates are $\sim 10^{-2} \msun$/ yr
 or larger to explain the peak luminosity \citep{Maeda2022ApJ...927...25M}. 
 In Table 2 and 3, we show the mass loss rates at $\log T_{\rm c}= 9.0$ (K)
 by $|\dot M_{9.0}|$. Typically around this stage, 
 the mass loss rate increases gradually with time toward the onset of pair-instabillity. 
 Therefore, it is expected that the mass loss rates in our models are
 not large enough to explain the peak luminosity in the SN light curve.

 To explain the observations in the late time interaction as in SN2018ibb,
 lower mass loss rates than $\sim 10^{-2} \msun$/ yr are sufficient.
 The brightness of the interaction depends on the CSM density and thus depends also on
 mass loss wind velocity as well as the mass-loss rate. Since we use the WR
 mass-loss rate by SV2020, we can make use of their estimates for the wind velocity,
 which is typically 1000-2000 km/s for our models. For this velocity, we can
 estimate that that the mass-loss rate at around $\log T_{\rm c}= 9.0$ is
 relevant to the interaction around several hundred days after the peak 
 (Nagele et al, in prep for more detail). 
 The tables show that $|\dot M_{9.0}|$ is typically $6~9 \times 10^{-4} \msun$/ yr
 for darker PISNe and $1 ~ 1.3 \times 10^{-3} \msun$/ yr for bright PISNe.
 Therefore, we expect that our PISNe models will have some signatures of
 CSM interaction. From these rough arguments we cannot definitively say that our models explain the CSM
 interaction of SN2018ibb, but we view the situation as promising.

\subsection{Pair-instability gap in the black hole mass distribution}

 Since the detection of gravitational waves from black hole merger \citep{Abbott2021PhRvX..11b1053A},
 the black hole mass function has undergone increased scrutiny.  
 To estimate the merger event rate, the existence of a gap in the BH mass 
 distribution due to the occurrence of PISNe, called the pair-instability mass gap is important.
 Previously, the gap range was mentioned as $65-130 \msun$ \citep[e.g. in][]{Farmer2019ApJ...887...53F,Mapelli2020ApJ...888...76M}, but since then several results
 have shown a larger value for the lower boundary. 
 For example, in our previous paper
 \citet{Umeda2020ApJ...905L..21U}, it is shown that the lower bound for Pop III stars
 depends on the overshooting parameter. It could be as large as 109 $\msun$
 for relatively small overshoot parameter, while this could be $\sim 66 \msun$
 if a larger value is assumed.

 The upper bound should also be model dependent and here we recall the values
 from \citet{takahashi2018} for Pop III PISNe. For their non-rotating 
 and magnetic-rotating models, the lowest
 BH masses above the PISN gap were $257 - 290 \msun$ for hydrogen stars,
 $139 - 150 \msun$ for helium cores, and $139 - 140 \msun$ for CO cores.
 Considering mass loss by binary interactions, the values for CO core are
 often mentioned as the upper bound for the gap, i.e., the upper
 bound is roughly $140 \msun$. 

 Let us now focus on the results of this paper. 
 We find that the maximum CO core mass for
 PISNe in LMC, SMC, and 0.1$Z_\odot$ models are $\simeq$ 109, 118, and 118 $\msun$,
 respectively (Tables 2 and 3). The BH masses above the PISN gap inferred from the Tables
 are 119.7, 116.7 $\msun$ for LMC, 111.0, 117.8 $\msun$ for SMC and
 119.2, 107.3, 115.5 $\msun$ for 0.1$Z_\odot$ models.
 From these results we may say that the upper bound for the gap is
 $107$ to $120 \msun$, which is significantly lower than the Pop III results. Furthermore, there are several models with CO core-masses inside the "gap", but with low entropy, which we expect to form black holes (Sec. \ref{sec:peculiar}).

 Combined with the results in \citet{Umeda2020ApJ...905L..21U} the PISN gap may not exist
 depending on the overshooting parameter if we consider both the Pop III and Pop II
 populations, although the gap may exist for a fixed metallicity.

\subsection{What initial rotation speed is reasonable ? }

  We have shown that with our choice of mass loss rates, PISNe in principle occur
  for LMC metallicity and lower. Is this consistent with observations
  in which no clear evidence for PISNe has yet been obtained?  We used the phrase 'in principle occur' because PISNe for LMC metallicity is possible only for 
  slowly rotating models, say  $v_{\rm i}/ v_{\rm k} \lesssim 0.1$.
  Even for $v_{\rm rot}/ v_{\rm crit} = 0.1$, the minimum mass to be a PISN is
  400 $\msun$ and larger mass is required for faster rotation. 

  In \citet{Georgy2012A&A...542A..29G}, it is discussed that the typical velocity of OB 
  stars are $v_{\rm i}/ v_{\rm k} \sim 0.4$. If we need to apply this value for
  our models, the mass range of PISNe becomes very high and PISNe will be extremely
  rare for LMC and SMC metallicities. However, there is basically no observations for the rotation velocities
  for such massive stars, and we currently do not know if $M > 300 \msun$ stars
  should initially rotate rapidly. If we find no evidence for PISNe   
  from galaxies with LMC and SMC metallicity, the
  initial rotation speed may explain the non-detection.

  For 0.1 $Z_\odot$ metallicity, on the other hand, PISNe are possible
  in a relatively low mass range, say $M \sim 250 \msun$,  
  even for fast rotating cases ($v_{\rm i}/ v_{\rm k} = 0.2$).
  We note, however, that bright PISNe with $^{56}$Ni $\sim 30 \msun$ such as SN2018ibb
  requires a larger initial mass, $\sim 400 \msun$, even for this metallicity.

 \section{Concluding Remarks}
\label{sec:summary}

 In this paper we revisited rotating metal-enriched PISN models.
 Previously, only a few such models had been calculated 
 in the literature and the general properties for these models were not clear.
 By calculating several models using the WR mass-loss rates by SV2020, 
 we intended to clarify, the properties of metal-enriched PISNe, such as
 mass ranges, produced $^{56}$Ni mass, and mass loss histories
 for LMC, SMC and 0.1$Z_\odot$ metallicities.
 Since we would like to know the precise values for the $^{56}$Ni mass, for example,
 we carefully performed hydrodynamical simulations for collapse and explosion
 with a sufficiently large nuclear
 reaction network, as well as stellar evolution calculations.

 In general, the PISNe mass range is higher for higher metallicity since mass-loss
 is larger. For non-rotating cases, we find that the PISN ranges are $320 - 475 \msun$,
 $300 - 390 \msun$, and $150 - 330 \msun$ for LMC, SMC and 0.1$Z_\odot$ metallicities,
 respectively. We note that the range for 0.1$Z_\odot$ is already close to that of Pop III
 SNe \citep{takahashi2018}. 

 The rotation effects which we found are interesting. For slow rotation, the PISNe range
 shifts downwards because the effect of increased core mass is more important than 
 increased mass loss rate which shifts the range higher. For SMC and lower metallicities,
 our $v_{\rm i}/ v_{\rm k}=0.1$ models show lower PISNe range than non rotating models.
 On the other hand, for the LMC model, mass loss effects are larger and the range shifts
 higher even for $v_{\rm i}/ v_{\rm k}=0.1$ (Table 1). For faster rotation,
 $v_{\rm i}/ v_{\rm k}=0.2$ which roughly corresponds to typical rotation speed for
 observed OB stars, the mass ranges are larger than non-rotating models.
 As a result, if a PISN progenitor were to rotate as fast as $v_{\rm i}/ v_{\rm k} \gtrsim 0.2$,
 the typical mass for bright PISNe with $^{56}$Ni mass larger than $10\msun$, would
 be more than $400\msun$ though the mass range is lower for lower metallicities.

 One of the significant differences with this work from previous similar works is that
 we have obtained O-rich progenitors for most of our models while previous studies obtained He-rich ones. 
 This difference mainly comes
 from the choice of the WR mass-loss rates, where we adopt the SV2020 rate.
 Since the WR mass-loss rate is uncertain especially for these very massive stars,
 these results should confront observations. In our case, a He-rich PISN occurs
 only for dimmer PISNe with metallicity $Z \lesssim 0.1 Z_\odot$. 
 It is encouraging that a recently identified PISN candiate SN2018ibb appears to be
 O-rich, but if we were to find bright He-rich PISNe in the future, then we would have to
 modify the WR mass loss rates.

\section*{Data Availability}

The data underlying this article will be shared on reasonable request to the corresponding author.

\bigskip \bigskip

\section*{Acknowledgements}

 We thank K. Maeda for useful discussions.
This study was supported in part by the Grant-in-Aid for the Scientific Research of Japan Society for the Promotion of Science (JSPS, No. JP21H01123).


\bibliography{bib}{}

\begin{thebibliography}{}
\expandafter\ifx\csname natexlab\endcsname\relax\def\natexlab#1{#1}\fi
\providecommand{\url}[1]{\href{#1}{#1}}
\providecommand{\dodoi}[1]{doi:~\href{http://doi.org/#1}{\nolinkurl{#1}}}
\providecommand{\doeprint}[1]{\href{http://ascl.net/#1}{\nolinkurl{http://ascl.net/#1}}}
\providecommand{\doarXiv}[1]{\href{https://arxiv.org/abs/#1}{\nolinkurl{https://arxiv.org/abs/#1}}}

\bibitem[{{Abbott} {et~al.}(2021){Abbott}, {Abbott}, {Abraham}, {Acernese},
  {Ackley}, {Adams}, {Adams}, {Adhikari}, {Adya}, {Affeldt}, {Agathos},
  {Agatsuma}, {Aggarwal}, {Aguiar}, {Aiello}, {Ain}, {Ajith}, {Akcay}, {Allen},
  {Allocca}, {Altin}, {Amato}, {Anand}, {Ananyeva}, {Anderson}, {Anderson},
  {Angelova}, {Ansoldi}, {Antelis}, {Antier}, {Appert}, {Arai}, {Araya},
  {Areeda}, {Ar{\`e}ne}, {Arnaud}, {Aronson}, {Arun}, {Asali}, {Ascenzi},
  {Ashton}, {Aston}, {Astone}, {Aubin}, {Aufmuth}, {AultONeal}, {Austin},
  {Avendano}, {Babak}, {Badaracco}, {Bader}, {Bae}, {Baer}, {Bagnasco},
  {Baird}, {Ball}, {Ballardin}, {Ballmer}, {Bals}, {Balsamo}, {Baltus},
  {Banagiri}, {Bankar}, {Bankar}, {Barayoga}, {Barbieri}, {Barish}, {Barker},
  {Barneo}, {Barnum}, {Barone}, {Barr}, {Barsotti}, {Barsuglia}, {Barta},
  {Bartlett}, {Bartos}, {Bassiri}, {Basti}, {Bawaj}, {Bayley}, {Bazzan},
  {Becher}, {B{\'e}csy}, {Bedakihale}, {Bejger}, {Belahcene}, {Beniwal},
  {Benjamin}, {Bennett}, {Bentley}, {Bergamin}, {Berger}, {Bergmann},
  {Bernuzzi}, {Berry}, {Bersanetti}, {Bertolini}, {Betzwieser}, {Bhandare},
  {Bhandari}, {Bhattacharjee}, {Bidler}, {Bilenko}, {Billingsley}, {Birney},
  {Birnholtz}, {Biscans}, {Bischi}, {Biscoveanu}, {Bisht}, {Bitossi},
  {Bizouard}, {Blackburn}, {Blackman}, {Blair}, {Blair}, {Blair}, {Blanch},
  {Bobba}, {Bode}, {Boer}, {Boetzel}, {Bogaert}, {Boldrini}, {Bondu},
  {Bonilla}, {Bonnand}, {Booker}, {Boom}, {Bork}, {Boschi}, {Bose},
  {Bossilkov}, {Boudart}, {Bouffanais}, {Bozzi}, {Bradaschia}, {Brady},
  {Bramley}, {Branchesi}, {Brau}, {Breschi}, {Briant}, {Briggs}, {Brighenti},
  {Brillet}, {Brinkmann}, {Brockill}, {Brooks}, {Brooks}, {Brown}, {Brunett},
  {Bruno}, {Bruntz}, {Buikema}, {Bulik}, {Bulten}, {Buonanno}, {Buscicchio},
  {Buskulic}, {Byer}, {Cabero}, {Cadonati}, {Caesar}, {Cagnoli}, {Cahillane},
  {Calder{\'o}n Bustillo}, {Callaghan}, {Callister}, {Calloni}, {Camp},
  {Canepa}, {Cannon}, {Cao}, {Cao}, {Carapella}, {Carbognani}, {Carney},
  {Carpinelli}, {Carullo}, {Carver}, {Casanueva Diaz}, {Casentini}, {Caudill},
  {Cavagli{\`a}}, {Cavalier}, {Cavalieri}, {Cella}, {Cerd{\'a}-Dur{\'a}n},
  {Cesarini}, {Chaibi}, {Chakravarti}, {Chan}, {Chan}, {Chandra}, {Chanial},
  {Chao}, {Charlton}, {Chase}, {Chassande-Mottin}, {Chatterjee},
  {Chattopadhyay}, {Chaturvedi}, {Chatziioannou}, {Chen}, {Chen}, {Chen},
  {Chen}, {Cheng}, {Cheong}, {Chia}, {Chiadini}, {Chierici}, {Chincarini},
  {Chiummo}, {Cho}, {Cho}, {Cho}, {Choate}, {Christensen}, {Chu}, {Chua},
  {Chung}, {Chung}, {Ciani}, {Ciecielag}, {Cie{\'s}lar}, {Cifaldi}, {Ciobanu},
  {Ciolfi}, {Cipriano}, {Cirone}, {Clara}, {Clark}, {Clark}, {Clarke},
  {Clearwater}, {Clesse}, {Cleva}, {Coccia}, {Cohadon}, {Cohen}, {Colleoni},
  {Collette}, {Collins}, {Colpi}, {Constancio}, {Conti}, {Cooper}, {Corban},
  {Corbitt}, {Cordero-Carri{\'o}n}, {Corezzi}, {Corley}, {Cornish}, {Corre},
  {Corsi}, {Cortese}, {Costa}, {Cotesta}, {Coughlin}, {Coughlin}, {Coulon},
  {Countryman}, {Cousins}, {Couvares}, {Covas}, {Coward}, {Cowart}, {Coyne},
  {Coyne}, {Creighton}, {Creighton}, {Croquette}, {Crowder}, {Cudell},
  {Cullen}, {Cumming}, {Cummings}, {Cunningham}, {Cuoco}, {Cury{\l}o},
  {Canton}, {D{\'a}lya}, {Dana}, {DaneshgaranBajastani}, {D'Angelo}, {Danila},
  {Danilishin}, {D'Antonio}, {Danzmann}, {Darsow-Fromm}, {Dasgupta}, {Datrier},
  {Dattilo}, {Dave}, {Davier}, {Davies}, {Davis}, {Daw}, {Dean}, {DeBra},
  {Deenadayalan}, {Degallaix}, {De Laurentis}, {Del{\'e}glise}, {Del Favero},
  {De Lillo}, {De Lillo}, {Del Pozzo}, {DeMarchi}, {De Matteis}, {D'Emilio},
  {Demos}, {Denker}, {Dent}, {Depasse}, {De Pietri}, {De Rosa}, {De Rossi},
  {DeSalvo}, {de Varona}, {Dhurandhar}, {D{\'\i}az}, {Diaz-Ortiz}, {Didio},
  {Dietrich}, {Di Fiore}, {DiFronzo}, {Di Giorgio}, {Di Giovanni}, {Di
  Giovanni}, {Di Girolamo}, {Di Lieto}, {Ding}, {Di Pace}, {Di Palma}, {Di
  Renzo}, {Divakarla}, {Dmitriev}, {Doctor}, {D'Onofrio}, {Donovan}, {Dooley},
  {Doravari}, {Dorrington}, {Downes}, {Drago}, {Driggers}, {Du}, {Ducoin},
  {Dupej}, {Durante}, {D'Urso}, {Duverne}, {Dwyer}, {Easter}, {Eddolls},
  {Edelman}, {Edo}, {Edy}, {Effler}, {Eichholz}, {Eikenberry}, {Eisenmann},
  {Eisenstein}, {Ejlli}, {Errico}, {Essick}, {Estell{\'e}s}, {Estevez},
  {Etienne}, {Etzel}, {Evans}, {Evans}, {Ewing}, {Fafone}, {Fair}, {Fairhurst},
  {Fan}, {Farah}, {Farinon}, {Farr}, {Farr}, {Fauchon-Jones}, {Favata}, {Fays},
  {Fazio}, {Feicht}, {Fejer}, {Feng}, {Fenyvesi}, {Ferguson},
  {Fernandez-Galiana}, {Ferrante}, {Ferreira}, {Fidecaro}, {Figura}, {Fiori},
  {Fiorucci}, {Fishbach}, {Fisher}, {Fishner}, {Fittipaldi}, {Fitz-Axen},
  {Fiumara}, {Flaminio}, {Floden}, {Flynn}, {Fong}, {Font}, {Forsyth},
  {Fournier}, {Frasca}, {Frasconi}, {Frei}, {Freise}, {Frey}, {Frey},
  {Fritschel}, {Frolov}, {Fronz{\'e}}, {Fulda}, {Fyffe}, {Gabbard}, {Gadre},
  {Gaebel}, {Gair}, {Gais}, {Galaudage}, {Gamba}, {Ganapathy}, {Ganguly},
  {Gaonkar}, {Garaventa}, {Garc{\'\i}a-Quir{\'o}s}, {Garufi}, {Gateley},
  {Gaudio}, {Gayathri}, {Gemme}, {Gennai}, {George}, {George}, {George},
  {Gergely}, {Ghonge}, {Ghosh}, {Ghosh}, {Ghosh}, {Giacomazzo}, {Giacoppo},
  {Giaime}, {Giardina}, {Gibson}, {Gier}, {Gill}, {Giri}, {Glanzer}, {Gleckl},
  {Godwin}, {Goetz}, {Goetz}, {Gohlke}, {Goncharov}, {Gonz{\'a}lez},
  {Gopakumar}, {Gossan}, {Gosselin}, {Gouaty}, {Grace}, {Grado}, {Granata},
  {Granata}, {Grant}, {Gras}, {Grassia}, {Gray}, {Gray}, {Greco}, {Green},
  {Green}, {Gretarsson}, {Griggs}, {Grignani}, {Grimaldi}, {Grimes}, {Grimm},
  {Grote}, {Grunewald}, {Gruning}, {Guerrero}, {Guidi}, {Guimaraes},
  {Guix{\'e}}, {Gulati}, {Guo}, {Gupta}, {Gupta}, {Gupta}, {Gustafson},
  {Gustafson}, {Guzman}, {Haegel}, {Halim}, {Hall}, {Hamilton}, {Hammond},
  {Haney}, {Hanke}, {Hanks}, {Hanna}, {Hannam}, {Hannuksela}, {Hannuksela},
  {Hansen}, {Hansen}, {Hanson}, {Harder}, {Hardwick}, {Haris}, {Harms},
  {Harry}, {Harry}, {Hartwig}, {Hasskew}, {Haster}, {Haughian}, {Hayes},
  {Healy}, {Heidmann}, {Heintze}, {Heinze}, {Heinzel}, {Heitmann}, {Hellman},
  {Hello}, {Helmling-Cornell}, {Hemming}, {Hendry}, {Heng}, {Hennes}, {Hennig},
  {Hennig}, {Hernandez Vivanco}, {Heurs}, {Hild}, {Hill}, {Hines}, {Hochheim},
  {Hofgard}, {Hofman}, {Hohmann}, {Holgado}, {Holland}, {Hollows}, {Holmes},
  {Holt}, {Holz}, {Hopkins}, {Horst}, {Hough}, {Howell}, {Hoy}, {Hoyland},
  {Huang}, {H{\"u}bner}, {Huddart}, {Huerta}, {Hughey}, {Hui}, {Husa},
  {Huttner}, {Hutzler}, {Huxford}, {Huynh-Dinh}, {Idzkowski}, {Iess},
  {Imperato}, {Inchauspe}, {Ingram}, {Intini}, {Isi}, {Iyer},
  {JaberianHamedan}, {Jacqmin}, {Jadhav}, {Jadhav}, {James}, {Jani},
  {Janssens}, {Janthalur}, {Jaranowski}, {Jariwala}, {Jaume}, {Jenkins},
  {Jeunon}, {Jiang}, {Johns}, {Johnson-McDaniel}, {Jones}, {Jones}, {Jones},
  {Jones}, {Jones}, {Jonker}, {Ju}, {Junker}, {Kalaghatgi}, {Kalogera},
  {Kamai}, {Kandhasamy}, {Kang}, {Kanner}, {Kapadia}, {Kapasi}, {Karathanasis},
  {Karki}, {Kashyap}, {Kasprzack}, {Kastaun}, {Katsanevas}, {Katsavounidis},
  {Katzman}, {Kawabe}, {K{\'e}f{\'e}lian}, {Keitel}, {Key}, {Khadka},
  {Khalili}, {Khan}, {Khan}, {Khazanov}, {Khetan}, {Khursheed}, {Kijbunchoo},
  {Kim}, {Kim}, {Kim}, {Kim}, {Kim}, {Kim}, {Kimball}, {King}, {Kinley-Hanlon},
  {Kirchhoff}, {Kissel}, {Kleybolte}, {Klimenko}, {Knowles}, {Knyazev}, {Koch},
  {Koehlenbeck}, {Koekoek}, {Koley}, {Kolstein}, {Komori}, {Kondrashov},
  {Kontos}, {Koper}, {Korobko}, {Korth}, {Kovalam}, {Kozak}, {Kr{\"a}mer},
  {Kringel}, {Krishnendu}, {Kr{\'o}lak}, {Kuehn}, {Kumar}, {Kumar}, {Kumar},
  {Kumar}, {Kuns}, {Kwang}, {Lackey}, {Laghi}, {Lalande}, {Lam}, {Lamberts},
  {Landry}, {Lane}, {Lang}, {Lange}, {Lantz}, {Lanza}, {La Rosa},
  {Lartaux-Vollard}, {Lasky}, {Laxen}, {Lazzarini}, {Lazzaro}, {Leaci},
  {Leavey}, {Lecoeuche}, {Lee}, {Lee}, {Lee}, {Lee}, {Lehmann}, {Leon},
  {Leroy}, {Letendre}, {Levin}, {Li}, {Li}, {Li}, {Li}, {Li}, {Linde},
  {Linker}, {Linley}, {Littenberg}, {Liu}, {Liu}, {Llorens-Monteagudo}, {Lo},
  {Lockwood}, {London}, {Longo}, {Lorenzini}, {Loriette}, {Lormand}, {Losurdo},
  {Lough}, {Lousto}, {Lovelace}, {L{\"u}ck}, {Lumaca}, {Lundgren}, {Ma},
  {Macas}, {MacInnis}, {Macleod}, {MacMillan}, {Macquet}, {Maga{\~n}a
  Hernandez}, {Maga{\~n}a-Sandoval}, {Magazz{\`u}}, {Magee}, {Majorana},
  {Maksimovic}, {Maliakal}, {Malik}, {Man}, {Mandic}, {Mangano}, {Mansell},
  {Manske}, {Mantovani}, {Mapelli}, {Marchesoni}, {Marion}, {M{\'a}rka},
  {M{\'a}rka}, {Markakis}, {Markosyan}, {Markowitz}, {Maros}, {Marquina},
  {Marsat}, {Martelli}, {Martin}, {Martin}, {Martinez}, {Martinez}, {Martynov},
  {Masalehdan}, {Mason}, {Massera}, {Masserot}, {Massinger}, {Masso-Reid},
  {Mastrogiovanni}, {Matas}, {Mateu-Lucena}, {Matichard}, {Matiushechkina},
  {Mavalvala}, {Maynard}, {McCann}, {McCarthy}, {McClelland}, {McCormick},
  {McCuller}, {McGuire}, {McIsaac}, {McIver}, {McManus}, {McRae}, {McWilliams},
  {Meacher}, {Meadors}, {Mehmet}, {Mehta}, {Melatos}, {Melchor}, {Mendell},
  {Menendez-Vazquez}, {Mercer}, {Mereni}, {Merfeld}, {Merilh}, {Merritt},
  {Merzougui}, {Meshkov}, {Messenger}, {Messick}, {Metzdorff}, {Meyers},
  {Meylahn}, {Mhaske}, {Miani}, {Miao}, {Michaloliakos}, {Michel}, {Middleton},
  {Milano}, {Miller}, {Millhouse}, {Mills}, {Milotti}, {Milovich-Goff},
  {Minazzoli}, {Minenkov}, {Mir}, {Mishkin}, {Mishra}, {Mistry}, {Mitra},
  {Mitrofanov}, {Mitselmakher}, {Mittleman}, {Mo}, {Mogushi}, {Mohapatra},
  {Mohite}, {Molina}, {Molina-Ruiz}, {Mondin}, {Montani}, {Moore}, {Moraru},
  {Morawski}, {Moreno}, {Morisaki}, {Mours}, {Mow-Lowry}, {Mozzon},
  {Muciaccia}, {Mukherjee}, {Mukherjee}, {Mukherjee}, {Mukherjee}, {Mukund},
  {Mullavey}, {Munch}, {Mu{\~n}iz}, {Murray}, {Nadji}, {Nagar}, {Nardecchia},
  {Naticchioni}, {Nayak}, {Neil}, {Neilson}, {Nelemans}, {Nelson}, {Nery},
  {Neunzert}, {Nitz}, {Ng}, {Ng}, {Nguyen}, {Nguyen}, {Nguyen}, {Nichols},
  {Nissanke}, {Nocera}, {Noh}, {North}, {Nothard}, {Nuttall}, {Oberling},
  {O'Brien}, {O'Dell}, {Oganesyan}, {Ogin}, {Oh}, {Oh}, {Ohme}, {Ohta},
  {Okada}, {Olivetto}, {Oppermann}, {Oram}, {O'Reilly}, {Ormiston}, {Ortega},
  {O'Shaughnessy}, {Ossokine}, {Osthelder}, {Ottaway}, {Overmier}, {Owen},
  {Pace}, {Pagano}, {Page}, {Pagliaroli}, {Pai}, {Pai}, {Palamos}, {Palashov},
  {Palomba}, {Pan}, {Panda}, {Pang}, {Pankow}, {Pannarale}, {Pant}, {Paoletti},
  {Paoli}, {Paolone}, {Parker}, {Pascucci}, {Pasqualetti}, {Passaquieti},
  {Passuello}, {Patel}, {Patricelli}, {Payne}, {Pechsiri}, {Pedraza},
  {Pegoraro}, {Pele}, {Penn}, {Perego}, {Perez}, {P{\'e}rigois}, {Perreca},
  {Perri{\`e}s}, {Petermann}, {Petterson}, {Pfeiffer}, {Pham}, {Phukon},
  {Piccinni}, {Pichot}, {Piendibene}, {Piergiovanni}, {Pierini}, {Pierro},
  {Pillant}, {Pilo}, {Pinard}, {Pinto}, {Piotrzkowski}, {Pirello}, {Pitkin},
  {Placidi}, {Plastino}, {Pluchar}, {Poggiani}, {Polini}, {Pong}, {Ponrathnam},
  {Popolizio}, {Porter}, {Poverman}, {Powell}, {Pracchia}, {Prajapati},
  {Prasai}, {Prasanna}, {Pratten}, {Prestegard}, {Principe}, {Prodi},
  {Prokhorov}, {Prosposito}, {Prudenzi}, {Puecher}, {Punturo}, {Puosi},
  {Puppo}, {P{\"u}rrer}, {Qi}, {Quetschke}, {Quinonez}, {Quitzow-James},
  {Raab}, {Raaijmakers}, {Radkins}, {Radulesco}, {Raffai}, {Rafferty}, {Rail},
  {Raja}, {Rajan}, {Rajbhandari}, {Rakhmanov}, {Ramirez}, {Ramirez},
  {Ramos-Buades}, {Rana}, {Rao}, {Rapagnani}, {Rapol}, {Ratto}, {Raymond},
  {Razzano}, {Read}, {Regimbau}, {Rei}, {Reid}, {Reitze}, {Rettegno}, {Ricci},
  {Richardson}, {Richardson}, {Richardson}, {Ricker}, {Riemenschneider},
  {Riles}, {Rizzo}, {Robertson}, {Robinet}, {Rocchi}, {Rocha}, {Rodriguez},
  {Rodriguez-Soto}, {Rolland}, {Rollins}, {Roma}, {Romanelli}, {Romano},
  {Romel}, {Romero}, {Romero-Shaw}, {Romie}, {Ronchini}, {Rose}, {Rose},
  {Rose}, {Rosell}, {Rosi{\'n}ska}, {Rosofsky}, {Ross}, {Rowan}, {Rowlinson},
  {Roy}, {Roy}, {Ruggi}, {Ryan}, {Sachdev}, {Sadecki}, {Sadiq},
  {Sakellariadou}, {Salafia}, {Salconi}, {Saleem}, {Samajdar}, {Sanchez},
  {Sanchez}, {Sanchez}, {Sanchis-Gual}, {Sanders}, {Sandles}, {Santiago},
  {Santos}, {Saravanan}, {Sarin}, {Sassolas}, {Sathyaprakash}, {Sauter},
  {Savage}, {Savant}, {Sawant}, {Sayah}, {Schaetzl}, {Schale}, {Scheel},
  {Scheuer}, {Schindler-Tyka}, {Schmidt}, {Schnabel}, {Schofield},
  {Sch{\"o}nbeck}, {Schreiber}, {Schulte}, {Schutz}, {Schwarm}, {Schwartz},
  {Scott}, {Scott}, {Seglar-Arroyo}, {Seidel}, {Sellers}, {Sengupta},
  {Sennett}, {Sentenac}, {Sequino}, {Sergeev}, {Setyawati}, {Shaffer},
  {Shahriar}, {Sharifi}, {Sharma}, {Sharma}, {Shawhan}, {Shen}, {Shikauchi},
  {Shink}, {Shoemaker}, {Shoemaker}, {Shukla}, {ShyamSundar}, {Sieniawska},
  {Sigg}, {Singer}, {Singh}, {Singh}, {Singha}, {Singhal}, {Sintes}, {Sipala},
  {Skliris}, {Slagmolen}, {Slaven-Blair}, {Smetana}, {Smith}, {Smith},
  {Somala}, {Son}, {Soni}, {Soni}, {Sorazu}, {Sordini}, {Sorrentino},
  {Sorrentino}, {Soulard}, {Souradeep}, {Sowell}, {Spencer}, {Spera},
  {Srivastava}, {Srivastava}, {Staats}, {Stachie}, {Steer}, {Steinhoff},
  {Steinke}, {Steinlechner}, {Steinlechner}, {Steinmeyer}, {Stevenson},
  {Stolle-McAllister}, {Stops}, {Stover}, {Strain}, {Stratta}, {Strunk},
  {Sturani}, {Stuver}, {S{\"u}dbeck}, {Sudhagar}, {Sudhir}, {Suh},
  {Summerscales}, {Sun}, {Sun}, {Sunil}, {Sur}, {Suresh}, {Sutton}, {Swinkels},
  {Szczepa{\'n}czyk}, {Tacca}, {Tait}, {Talbot}, {Tanasijczuk}, {Tanner},
  {Tao}, {Tapia}, {Tapia San Martin}, {Tasson}, {Taylor}, {Tenorio},
  {Terkowski}, {Thirugnanasambandam}, {Thomas}, {Thomas}, {Thomas}, {Thompson},
  {Thondapu}, {Thorne}, {Thrane}, {Tiwari}, {Tiwari}, {Tiwari}, {Toland},
  {Tolley}, {Tonelli}, {Tornasi}, {Torres-Forn{\'e}}, {Torrie}, {e Melo},
  {T{\"o}yr{\"a}}, {Tran}, {Trapananti}, {Travasso}, {Traylor}, {Tringali},
  {Tripathee}, {Trovato}, {Trudeau}, {Tsai}, {Tsang}, {Tse}, {Tso}, {Tsukada},
  {Tsuna}, {Tsutsui}, {Turconi}, {Ubhi}, {Udall}, {Ueno}, {Ugolini},
  {Unnikrishnan}, {Urban}, {Usman}, {Utina}, {Vahlbruch}, {Vajente}, {Vajpeyi},
  {Valdes}, {Valentini}, {Valsan}, {van Bakel}, {van Beuzekom}, {van den
  Brand}, {Van Den Broeck}, {Vander-Hyde}, {van der Schaaf}, {van Heijningen},
  {Vardaro}, {Vargas}, {Varma}, {Vass}, {Vas{\'u}th}, {Vecchio}, {Vedovato},
  {Veitch}, {Veitch}, {Venkateswara}, {Venneberg}, {Venugopalan}, {Verkindt},
  {Verma}, {Veske}, {Vetrano}, {Vicer{\'e}}, {Viets}, {Vijaykumar},
  {Villa-Ortega}, {Vinet}, {Vitale}, {Vo}, {Vocca}, {Vorvick}, {Vyatchanin},
  {Wade}, {Wade}, {Wade}, {Walet}, {Walker}, {Wallace}, {Wallace}, {Walsh},
  {Wang}, {Wang}, {Wang}, {Wang}, {Ward}, {Warner}, {Was}, {Washington},
  {Watchi}, {Weaver}, {Wei}, {Weinert}, {Weinstein}, {Weiss}, {Wellmann},
  {Wen}, {We{\ss}els}, {Westhouse}, {Wette}, {Whelan}, {White}, {White},
  {Whiting}, {Whittle}, {Wilken}, {Williams}, {Williams}, {Williamson},
  {Willis}, {Willke}, {Wilson}, {Wimmer}, {Winkler}, {Wipf}, {Woan}, {Woehler},
  {Wofford}, {Wong}, {Wrangel}, {Wright}, {Wu}, {Wysocki}, {Xiao}, {Yamamoto},
  {Yang}, {Yang}, {Yang}, {Yap}, {Yeeles}, {Yoon}, {Yu}, {Yu}, {Yuen},
  {Zadro{\.Z}ny}, {Zanolin}, {Zelenova}, {Zendri}, {Zevin}, {Zhang}, {Zhang},
  {Zhang}, {Zhang}, {Zhao}, {Zhao}, {Zheng}, {Zhou}, {Zhou}, {Zhu},
  {Zimmerman}, {Zlochower}, {Zucker}, {Zweizig}, {LIGO Scientific
  Collaboration}, \& {Virgo Collaboration}}]{Abbott2021PhRvX..11b1053A}
{Abbott}, R., {Abbott}, T.~D., {Abraham}, S., {et~al.} 2021, Physical Review X,
  11, 021053, \dodoi{10.1103/PhysRevX.11.021053}

\bibitem[{{Aguilera-Dena} {et~al.}(2020){Aguilera-Dena}, {Langer},
  {Antoniadis}, \& {M{\"u}ller}}]{Aguilera-Dena2020ApJ...901..114A}
{Aguilera-Dena}, D.~R., {Langer}, N., {Antoniadis}, J., \& {M{\"u}ller}, B.
  2020, \apj, 901, 114, \dodoi{10.3847/1538-4357/abb138}

\bibitem[{{Aguilera-Dena} {et~al.}(2018){Aguilera-Dena}, {Langer}, {Moriya}, \&
  {Schootemeijer}}]{Aguilera-Dena2018ApJ...858..115A}
{Aguilera-Dena}, D.~R., {Langer}, N., {Moriya}, T.~J., \& {Schootemeijer}, A.
  2018, \apj, 858, 115, \dodoi{10.3847/1538-4357/aabfc1}

\bibitem[{{Aoki} {et~al.}(2014){Aoki}, {Tominaga}, {Beers}, {Honda}, \&
  {Lee}}]{Aoki2014Sci...345..912A}
{Aoki}, W., {Tominaga}, N., {Beers}, T.~C., {Honda}, S., \& {Lee}, Y.~S. 2014,
  Science, 345, 912, \dodoi{10.1126/science.1252633}

\bibitem[{{Asplund} {et~al.}(2009){Asplund}, {Grevesse}, {Sauval}, \&
  {Scott}}]{asplund2009}
{Asplund}, M., {Grevesse}, N., {Sauval}, A.~J., \& {Scott}, P. 2009, \araa, 47,
  481, \dodoi{10.1146/annurev.astro.46.060407.145222}

\bibitem[{{Bardeen} {et~al.}(1972){Bardeen}, {Press}, \&
  {Teukolsky}}]{Bardeen1972ApJ...178..347B}
{Bardeen}, J.~M., {Press}, W.~H., \& {Teukolsky}, S.~A. 1972, \apj, 178, 347,
  \dodoi{10.1086/151796}

\bibitem[{{Barkat} {et~al.}(1967){Barkat}, {Rakavy}, \&
  {Sack}}]{Barkat1967PhRvL..18..379B}
{Barkat}, Z., {Rakavy}, G., \& {Sack}, N. 1967, \prl, 18, 379,
  \dodoi{10.1103/PhysRevLett.18.379}

\bibitem[{{Caughlan} \& {Fowler}(1988)}]{caughlan1988}
{Caughlan}, G.~R., \& {Fowler}, W.~A. 1988, Atomic Data and Nuclear Data
  Tables, 40, 283, \dodoi{10.1016/0092-640X(88)90009-5}

\bibitem[{{Crowther}(2000)}]{Crowther2000A&A...356..191C}
{Crowther}, P.~A. 2000, \aap, 356, 191, \dodoi{10.48550/arXiv.astro-ph/0001226}

\bibitem[{{Cyburt} {et~al.}(2010){Cyburt}, {Amthor}, {Ferguson}, {Meisel},
  {Smith}, {Warren}, {Heger}, {Hoffman}, {Rauscher}, {Sakharuk}, {Schatz},
  {Thielemann}, \& {Wiescher}}]{Cyburt2010ApJS..189..240C}
{Cyburt}, R.~H., {Amthor}, A.~M., {Ferguson}, R., {et~al.} 2010, \apjs, 189,
  240, \dodoi{10.1088/0067-0049/189/1/240}

\bibitem[{{de Jager} {et~al.}(1988){de Jager}, {Nieuwenhuijzen}, \& {van der
  Hucht}}]{deJager1988A&AS...72..259D}
{de Jager}, C., {Nieuwenhuijzen}, H., \& {van der Hucht}, K.~A. 1988, \aaps,
  72, 259

\bibitem[{{Dessart} {et~al.}(2012){Dessart}, {Hillier}, {Waldman}, {Livne}, \&
  {Blondin}}]{Dessart2012MNRAS.426L..76D}
{Dessart}, L., {Hillier}, D.~J., {Waldman}, R., {Livne}, E., \& {Blondin}, S.
  2012, \mnras, 426, L76, \dodoi{10.1111/j.1745-3933.2012.01329.x}

\bibitem[{{Eggenberger} {et~al.}(2008){Eggenberger}, {Meynet}, {Maeder},
  {Hirschi}, {Charbonnel}, {Talon}, \&
  {Ekstr{\"o}m}}]{Eggenberger2008Ap&SS.316...43E}
{Eggenberger}, P., {Meynet}, G., {Maeder}, A., {et~al.} 2008, \apss, 316, 43,
  \dodoi{10.1007/s10509-007-9511-y}

\bibitem[{{Ekstr{\"o}m} {et~al.}(2012){Ekstr{\"o}m}, {Georgy}, {Eggenberger},
  {Meynet}, {Mowlavi}, {Wyttenbach}, {Granada}, {Decressin}, {Hirschi},
  {Frischknecht}, {Charbonnel}, \& {Maeder}}]{Ekstrom2012A&A...537A.146E}
{Ekstr{\"o}m}, S., {Georgy}, C., {Eggenberger}, P., {et~al.} 2012, \aap, 537,
  A146, \dodoi{10.1051/0004-6361/201117751}

\bibitem[{{Farmer} {et~al.}(2019){Farmer}, {Renzo}, {de Mink}, {Marchant}, \&
  {Justham}}]{Farmer2019ApJ...887...53F}
{Farmer}, R., {Renzo}, M., {de Mink}, S.~E., {Marchant}, P., \& {Justham}, S.
  2019, \apj, 887, 53, \dodoi{10.3847/1538-4357/ab518b}

\bibitem[{{Gal-Yam}(2019)}]{Gal-Yam2019ARA&A..57..305G}
{Gal-Yam}, A. 2019, \araa, 57, 305, \dodoi{10.1146/annurev-astro-081817-051819}

\bibitem[{{Gal-Yam} {et~al.}(2009){Gal-Yam}, {Mazzali}, {Ofek}, {Nugent},
  {Kulkarni}, {Kasliwal}, {Quimby}, {Filippenko}, {Cenko}, {Chornock},
  {Waldman}, {Kasen}, {Sullivan}, {Beshore}, {Drake}, {Thomas}, {Bloom},
  {Poznanski}, {Miller}, {Foley}, {Silverman}, {Arcavi}, {Ellis}, \&
  {Deng}}]{Gal-Yam2009Natur.462..624G}
{Gal-Yam}, A., {Mazzali}, P., {Ofek}, E.~O., {et~al.} 2009, \nat, 462, 624,
  \dodoi{10.1038/nature08579}

\bibitem[{{Georgy} {et~al.}(2012){Georgy}, {Ekstr{\"o}m}, {Meynet}, {Massey},
  {Levesque}, {Hirschi}, {Eggenberger}, \&
  {Maeder}}]{Georgy2012A&A...542A..29G}
{Georgy}, C., {Ekstr{\"o}m}, S., {Meynet}, G., {et~al.} 2012, \aap, 542, A29,
  \dodoi{10.1051/0004-6361/201118340}

\bibitem[{{Heger} {et~al.}(2000){Heger}, {Langer}, \& {Woosley}}]{heger2000}
{Heger}, A., {Langer}, N., \& {Woosley}, S.~E. 2000, \apj, 528, 368,
  \dodoi{10.1086/308158}

\bibitem[{{Heger} \& {Woosley}(2002)}]{Heger2002ApJ...567..532H}
{Heger}, A., \& {Woosley}, S.~E. 2002, \apj, 567, 532, \dodoi{10.1086/338487}

\bibitem[{{Heger} {et~al.}(2005){Heger}, {Woosley}, \& {Spruit}}]{heger2005}
{Heger}, A., {Woosley}, S.~E., \& {Spruit}, H.~C. 2005, \apj, 626, 350,
  \dodoi{10.1086/429868}

\bibitem[{{Hirschi} {et~al.}(2004){Hirschi}, {Meynet}, \&
  {Maeder}}]{Hirschi2004A&A...425..649H}
{Hirschi}, R., {Meynet}, G., \& {Maeder}, A. 2004, \aap, 425, 649,
  \dodoi{10.1051/0004-6361:20041095}

\bibitem[{{Jerkstrand} {et~al.}(2017){Jerkstrand}, {Smartt}, {Inserra},
  {Nicholl}, {Chen}, {Kr{\"u}hler}, {Sollerman}, {Taubenberger}, {Gal-Yam},
  {Kankare}, {Maguire}, {Fraser}, {Valenti}, {Sullivan}, {Cartier}, \&
  {Young}}]{Jerkstrand2017ApJ...835...13J}
{Jerkstrand}, A., {Smartt}, S.~J., {Inserra}, C., {et~al.} 2017, \apj, 835, 13,
  \dodoi{10.3847/1538-4357/835/1/13}

\bibitem[{{Kippenhahn} \& {Weigert}(1990)}]{Kippenhahn1990sse..book.....K}
{Kippenhahn}, R., \& {Weigert}, A. 1990, {Stellar Structure and Evolution}
  (Springer-Verlag)

\bibitem[{{Kozyreva} {et~al.}(2014){Kozyreva}, {Yoon}, \&
  {Langer}}]{Kozyreva2014A&A...566A.146K}
{Kozyreva}, A., {Yoon}, S.~C., \& {Langer}, N. 2014, \aap, 566, A146,
  \dodoi{10.1051/0004-6361/201423641}

\bibitem[{{Langer}(1998)}]{Langer1998A&A...329..551L}
{Langer}, N. 1998, \aap, 329, 551

\bibitem[{{Langer} {et~al.}(2007){Langer}, {Norman}, {de Koter}, {Vink},
  {Cantiello}, \& {Yoon}}]{Langer2007A&A...475L..19L}
{Langer}, N., {Norman}, C.~A., {de Koter}, A., {et~al.} 2007, \aap, 475, L19,
  \dodoi{10.1051/0004-6361:20078482}

\bibitem[{{Luo} {et~al.}(2022){Luo}, {Umeda}, {Yoshida}, \&
  {Takahashi}}]{Luo2022ApJ...927..115L}
{Luo}, T., {Umeda}, H., {Yoshida}, T., \& {Takahashi}, K. 2022, \apj, 927, 115,
  \dodoi{10.3847/1538-4357/ac4f5f}

\bibitem[{{Maeda} \& {Moriya}(2022)}]{Maeda2022ApJ...927...25M}
{Maeda}, K., \& {Moriya}, T.~J. 2022, \apj, 927, 25,
  \dodoi{10.3847/1538-4357/ac4672}

\bibitem[{{Maeder} \& {Meynet}(2000)}]{Maeder2000A&A...361..159M}
{Maeder}, A., \& {Meynet}, G. 2000, \aap, 361, 159,
  \dodoi{10.48550/arXiv.astro-ph/0006405}

\bibitem[{{Mapelli} {et~al.}(2020){Mapelli}, {Spera}, {Montanari}, {Limongi},
  {Chieffi}, {Giacobbo}, {Bressan}, \&
  {Bouffanais}}]{Mapelli2020ApJ...888...76M}
{Mapelli}, M., {Spera}, M., {Montanari}, E., {et~al.} 2020, \apj, 888, 76,
  \dodoi{10.3847/1538-4357/ab584d}

\bibitem[{{Nagele} {et~al.}(2023){Nagele}, {Umeda}, \&
  {Takahashi}}]{Nagele2023arXiv230101941N}
{Nagele}, C., {Umeda}, H., \& {Takahashi}, K. 2023, arXiv e-prints,
  arXiv:2301.01941, \dodoi{10.48550/arXiv.2301.01941}

\bibitem[{{Nagele} {et~al.}(2022){Nagele}, {Umeda}, {Takahashi}, {Yoshida}, \&
  {Sumiyoshi}}]{Nagele2022arXiv220510493N}
{Nagele}, C., {Umeda}, H., {Takahashi}, K., {Yoshida}, T., \& {Sumiyoshi}, K.
  2022, \mnras, 517, 1584, \dodoi{10.1093/mnras/stac2495}

\bibitem[{{Nugis} \& {Lamers}(2000)}]{Nugis2000A&A...360..227N}
{Nugis}, T., \& {Lamers}, H.~J.~G.~L.~M. 2000, \aap, 360, 227

\bibitem[{{Rakavy} \& {Shaviv}(1967)}]{Rakavy1967ApJ...148..803R}
{Rakavy}, G., \& {Shaviv}, G. 1967, \apj, 148, 803, \dodoi{10.1086/149204}

\bibitem[{{Sander} \& {Vink}(2020)}]{Sander2020MNRAS.499..873S}
{Sander}, A. A.~C., \& {Vink}, J.~S. 2020, \mnras, 499, 873,
  \dodoi{10.1093/mnras/staa2712}

\bibitem[{{Schulze} {et~al.}(2023){Schulze}, {Fransson}, {Kozyreva}, {Chen},
  {Yaron}, {Jerkstrand}, {Gal-Yam}, {Sollerman}, {Yan}, {Kangas}, {Leloudas},
  {Omand}, {Smartt}, {Yang}, {Nicholl}, {Sarin}, {Yao}, {Brink}, {Sharon},
  {Rossi}, {Chen}, {Chen}, {Cikota}, {De}, {Drake}, {Filippenko}, {Fremling},
  {Freour}, {Fynbo}, {Ho}, {Inserra}, {Irani}, {Kuncarayakti}, {Lunnan},
  {Mazzali}, {Ofek}, {Palazzi}, {Perley}, {Pursiainen}, {Rothberg}, {Shingles},
  {Smith}, {Taggart}, {Tartaglia}, {Zheng}, {Anderson}, {Cassara},
  {Christensen}, {Djorgovski}, {Galbany}, {Gkini}, {Graham}, {Gromadzki},
  {Groom}, {Hiramatsu}, {Howell}, {Kasliwal}, {McCully}, {M{\"u}ller-Bravo},
  {Paiano}, {Paraskeva}, {Pessi}, {Polishook}, {Rau}, {Rigault}, \&
  {Rusholme}}]{Schulze2023arXiv230505796S}
{Schulze}, S., {Fransson}, C., {Kozyreva}, A., {et~al.} 2023, arXiv e-prints,
  arXiv:2305.05796, \dodoi{10.48550/arXiv.2305.05796}

\bibitem[{{Spruit}(2002)}]{Spruit2002A&A...381..923S}
{Spruit}, H.~C. 2002, \aap, 381, 923, \dodoi{10.1051/0004-6361:20011465}

\bibitem[{{Sumiyoshi} {et~al.}(2005){Sumiyoshi}, {Yamada}, {Suzuki}, {Shen},
  {Chiba}, \& {Toki}}]{sumiyoshi2005}
{Sumiyoshi}, K., {Yamada}, S., {Suzuki}, H., {et~al.} 2005, \apj, 629, 922,
  \dodoi{10.1086/431788}

\bibitem[{{Sylvester} {et~al.}(1998){Sylvester}, {Skinner}, \&
  {Barlow}}]{Sylvester1998MNRAS.301.1083S}
{Sylvester}, R.~J., {Skinner}, C.~J., \& {Barlow}, M.~J. 1998, \mnras, 301,
  1083, \dodoi{10.1046/j.1365-8711.1998.02078.x}

\bibitem[{{Takahashi} {et~al.}(2014){Takahashi}, {Umeda}, \&
  {Yoshida}}]{Takahashi2014ApJ...794...40T}
{Takahashi}, K., {Umeda}, H., \& {Yoshida}, T. 2014, \apj, 794, 40,
  \dodoi{10.1088/0004-637X/794/1/40}

\bibitem[{{Takahashi} {et~al.}(2018){Takahashi}, {Yoshida}, \&
  {Umeda}}]{takahashi2018}
{Takahashi}, K., {Yoshida}, T., \& {Umeda}, H. 2018, \apj, 857, 111,
  \dodoi{10.3847/1538-4357/aab95f}

\bibitem[{{Takahashi} {et~al.}(2016){Takahashi}, {Yoshida}, {Umeda},
  {Sumiyoshi}, \& {Yamada}}]{takahashi2016}
{Takahashi}, K., {Yoshida}, T., {Umeda}, H., {Sumiyoshi}, K., \& {Yamada}, S.
  2016, \mnras, 456, 1320, \dodoi{10.1093/mnras/stv2649}

\bibitem[{{Umeda} \& {Nomoto}(2002)}]{Umeda2002ApJ...565..385U}
{Umeda}, H., \& {Nomoto}, K. 2002, \apj, 565, 385, \dodoi{10.1086/323946}

\bibitem[{{Umeda} \& {Nomoto}(2005)}]{Umeda2005ApJ...619..427U}
---. 2005, \apj, 619, 427, \dodoi{10.1086/426097}

\bibitem[{{Umeda} {et~al.}(2020){Umeda}, {Yoshida}, {Nagele}, \&
  {Takahashi}}]{Umeda2020ApJ...905L..21U}
{Umeda}, H., {Yoshida}, T., {Nagele}, C., \& {Takahashi}, K. 2020, \apjl, 905,
  L21, \dodoi{10.3847/2041-8213/abcb96}

\bibitem[{{van Loon} {et~al.}(1999){van Loon}, {Groenewegen}, {de Koter},
  {Trams}, {Waters}, {Zijlstra}, {Whitelock}, \&
  {Loup}}]{vanLoon1999A&A...351..559V}
{van Loon}, J.~T., {Groenewegen}, M.~A.~T., {de Koter}, A., {et~al.} 1999,
  \aap, 351, 559, \dodoi{10.48550/arXiv.astro-ph/9909416}

\bibitem[{{Vink} {et~al.}(2001){Vink}, {de Koter}, \&
  {Lamers}}]{Vink2001A&A...369..574V}
{Vink}, J.~S., {de Koter}, A., \& {Lamers}, H.~J.~G.~L.~M. 2001, \aap, 369,
  574, \dodoi{10.1051/0004-6361:20010127}

\bibitem[{{Whalen} {et~al.}(2014){Whalen}, {Smidt}, {Heger}, {Hirschi},
  {Yusof}, {Even}, {Fryer}, {Stiavelli}, {Chen}, \&
  {Joggerst}}]{Whalen2014ApJ...797....9W}
{Whalen}, D.~J., {Smidt}, J., {Heger}, A., {et~al.} 2014, \apj, 797, 9,
  \dodoi{10.1088/0004-637X/797/1/9}

\bibitem[{{Woosley} \& {Heger}(2006)}]{Woosley2006ApJ...637..914W}
{Woosley}, S.~E., \& {Heger}, A. 2006, \apj, 637, 914, \dodoi{10.1086/498500}

\bibitem[{{Xing} {et~al.}(2023){Xing}, {Zhao}, {Liu}, {Heger}, {Han}, {Aoki},
  {Chen}, {Ishigaki}, {Li}, \& {Zhao}}]{Xing2023Nature}
{Xing}, Q.~F., {Zhao}, G., {Liu}, Z.~W., {et~al.} 2023, Nature, 618, 712–715,
  \dodoi{10.1038/s41586-023-06028-1}

\bibitem[{{Yamada}(1997)}]{yamada1997}
{Yamada}, S. 1997, \apj, 475, 720, \dodoi{10.1086/303548}

\bibitem[{{Yoon}(2017)}]{Yoon2017MNRAS.470.3970Y}
{Yoon}, S.-C. 2017, \mnras, 470, 3970, \dodoi{10.1093/mnras/stx1496}

\bibitem[{{Yoon} {et~al.}(2012){Yoon}, {Dierks}, \&
  {Langer}}]{Yoon2012A&A...542A.113Y}
{Yoon}, S.~C., {Dierks}, A., \& {Langer}, N. 2012, \aap, 542, A113,
  \dodoi{10.1051/0004-6361/201117769}

\bibitem[{{Yoon} \& {Langer}(2005)}]{Yoon2005A&A...443..643Y}
{Yoon}, S.~C., \& {Langer}, N. 2005, \aap, 443, 643,
  \dodoi{10.1051/0004-6361:20054030}

\bibitem[{{Yoon} {et~al.}(2006){Yoon}, {Langer}, \&
  {Norman}}]{Yoon2006A&A...460..199Y}
{Yoon}, S.~C., {Langer}, N., \& {Norman}, C. 2006, \aap, 460, 199,
  \dodoi{10.1051/0004-6361:20065912}

\bibitem[{{Yoshida} {et~al.}(2019){Yoshida}, {Takiwaki}, {Kotake}, {Takahashi},
  {Nakamura}, \& {Umeda}}]{Yoshida2019}
{Yoshida}, T., {Takiwaki}, T., {Kotake}, K., {et~al.} 2019, \apj, 881, 16,
  \dodoi{10.3847/1538-4357/ab2b9d}

\bibitem[{{Yoshida} \& {Umeda}(2011)}]{Yoshida2011MNRAS.412L..78Y}
{Yoshida}, T., \& {Umeda}, H. 2011, \mnras, 412, L78,
  \dodoi{10.1111/j.1745-3933.2011.01008.x}

\bibitem[{{Yusof} {et~al.}(2013){Yusof}, {Hirschi}, {Meynet}, {Crowther},
  {Ekstr{\"o}m}, {Frischknecht}, {Georgy}, {Abu Kassim}, \&
  {Schnurr}}]{Yusof2013MNRAS.433.1114Y}
{Yusof}, N., {Hirschi}, R., {Meynet}, G., {et~al.} 2013, \mnras, 433, 1114,
  \dodoi{10.1093/mnras/stt794}

\end{thebibliography}
\bibliographystyle{aasjournal}



\end{document}